\documentclass{ws-ijmpa}
\usepackage[super,compress]{cite}
\usepackage{graphicx}
\usepackage{float} %
\begin{document}
\markboth{Mahary Vasihoun}
{Gravitational and topological effects on $\sqrt{-F^2}$ Confinement Dynamics}

%
\catchline{}{}{}{}{}
%

\title{Gravitational and topological effects on $\sqrt{-F^2}$ Confinement Dynamics
}

\author{Mahary Vasihoun}

\address{Physics Department, Ben Gurion University, Beer Sheva, Israel\\
mahary@bgu.ac.il}

\author{Eduardo Guendelman}

\address{Physics Department, Ben Gurion University, Beer Sheva, Israel\\
guendel@bgu.ac.il}

\maketitle

\begin{history}
\received{Day Month Year}
\revised{Day Month Year}
\end{history}

\begin{abstract}

We present a review, of recent developments on non-linear gauge theory containing a $\sqrt{-F^2}$ term coupled to gravity. We start by showing some of the confining features of this theory in flat space-time. We then consider the coupling, of this non-linear term, to gravity and discuss two types of spherically symmetric solutions. One of them has a tube topology, that is $\mathcal{M}_2 \times S^2$, or of the Levi-Civitta-Robinson-Berttoti (LCRB) type, where the metric coefficient $g_{\theta \theta}$ is a constant. The other type of solutions, Reissner-Nordstr{\"o}m-de-Sitter (RNdS), with $g_{\theta \theta}= r^2$ , where $r$ is a radial variable allowed to have all values from zero to infinity. Next we consider the matching of these solutions via light-like, and subsequently, time-like membranes and show the topologically induced effects of ``Hiding of Charge", where a charged particle can appear neutral to an external observer looking at it from the RNdS region, and the ``Confining of charge" in a wormhole throat, where two opposite charges are at the opposite sides of a wormhole throats. We proceed with some applications to extended theories of general relativity, in the form of quadratic gravity model ($F(R)$), then wormholes arise naturally from the non-linear electromagnetic field rather than requiring exotic matter to generate a pre-designed wormhole geometry (Morris-Thorne approach), in another model considered here we have, in addition to quadratic gravity, a dilaton filed $(\phi$), where we find wormhole solutions with de-Sitter asymptotics, and {\em confinement-deconfinement} transition effects as function of the dilaton vacuum expectation value. The last application we present is to the ``Two Measure theory", where in addition to the metric volume element, $\sqrt{-g}$, we consider a new, metric independent, volume element $\Phi$.
Finally we conclude and summarize our findings.

\end{abstract}

\ccode{PACS numbers:}


\newpage
\section{Introduction}\label{intro}
There are various reasons supporting the natural appearance of the ``square-root'' Maxwell term in effective gauge field actions. Originally, a pure ``square-root'' Lagrangian in flat space-time 
\begin{equation}
- \frac{f}{2}\sqrt{F^2}
\label{eq:NandO}
\end{equation} 
was proposed by Nielsen and Olesen \cite{N-O-1}{,} such ``square root gauge field theory" has been shown to give rise to string solutions and therefore provides a theory which includes strings as special types of excitations, while containing other types of excitations as well \cite{N-O-2,N-O-3,N-O-4}{.} We notice, however, that this action is designed to be applicable only to "magnetic dominated" configuration, for electric dominated configuration the square root becomes imaginary.  

Furthermore, G. `t Hooft \cite{tHooft-03} has shown that in any effective quantum gauge field theory, which is able to describe a linear confinement phenomena, the energy density
of electrostatic field configurations should be a linear function of the
electric displacement field in the infrared region (the latter appearing as an ``infrared counterterm"). A simple way to realize this idea
in Minkowski space-time is by considering a non-linear effective gauge field model, which has an additional term of the form of square root of the standard Maxwell Lagrangian in the effective action\footnote{Although we consider an Abelian case, one could start with the non-Abelian version, since we will be interested in static 
spherically symmetric solutions, the non-Abelian gauge theory effectively reduces 
to an Abelian one}{.} 
\begin{equation}
\begin{alignedat}{2}
&S = \int L(F^2)d^4 x  \quad ,\quad
L(F^2) = -\frac{1}{4} F^2 - \frac{f}{2} \sqrt{-F^2}
\\
&F^2 \equiv F_{\mu\nu} F^{\mu\nu} \quad , \quad  F_{\mu\nu}= \partial_\mu A_{\nu} -  \partial_\nu A_{\mu} \quad , \quad f>0
\end{alignedat}
\label{eq:GG}
\end{equation}
This Lagrangian contains both the usual Maxwell term and a non-analytic function of $F^2$, it is thus a \emph{non-standard} form of non-linear electrodynamics, which is significantly different from the original purely ``square root'' Lagrangian (\ref{eq:NandO}) first proposed by Nielsen and Olesen. The variation of this action with respect to $A_\mu$ field produces the following equation
\begin{equation}
\frac{\partial }{{\partial x^\mu  }}\left[ {\left(
{\sqrt { - F_{\alpha \beta } F^{\alpha \beta } }  + f}
\right)\frac{{F^{\mu \nu } }}{{\sqrt { - F_{\alpha \beta }
F^{\alpha \beta } } }}} \right] = 0
 \label{eq:sib9}
\end{equation}
In order to illustrate that this Lagrangian satisfies `t Hooft criteria for theories describing confinement, we will study equation
(\ref{eq:sib9}) for the case of a spherically symmetric electric
field 
\begin{equation}
F_{0i}=-E_i \quad , \quad F_{ij}=0 \quad , \quad {\bf E}=E(r)\hat {\bf r}
\end{equation} 
Equation (\ref{eq:sib9}) is, then, solved by a scalar potential $V$ of the form
\begin{equation}
V  =\frac{q}{r}  - \frac{f}{{\sqrt 2 }}r 
 \label{eq:sib13}
\end{equation}
This has a Coulomb part plus a linear contribution which is of the form of the well-known ``Cornell'' potential \cite{cornell-potential-1} which was postulated in order to simulate the features of quantum chromodynamics (QCD). Also, for static field configurations the model (\ref{eq:GG}) has the following energy density
\begin{equation}
\vec{E}^2 = |\vec{D}|^2 +\sqrt{2}f|\vec{D}| + \dfrac{1}{2} f^2
\end{equation}
This contains a term linear with respect to $|\vec{D}|$ as argued by `t Hooft. 

The square root of the Maxwell term can also naturally arise, in flat space-time, as a result of spontaneous breakdown of scale symmetry of the original scale-invariant Maxwell 
theory with $f$ appearing as an integration constant responsible for the latter 
spontaneous breakdown\cite{GG-2,eg1Guendelman,eg1Guendelman-Gaete1,eg1Korover,eg1Korover1}{,} to show this we will start with a theory which is invariant under scale symmetry
\begin{equation}
S = \int {d^4 } x\left( { - \frac{1}{4}F_{\mu \nu }^{a} F^{a\mu \nu } }
\right) \quad , \quad  F_{\mu \nu }^a  = \partial _\mu  A_\nu
^a - \partial _\nu  A_\mu ^a  + gf^{abc} A_\mu ^b A_\nu ^c
 \label{eq:sib1}
\end{equation}
Another way to write (\ref{eq:sib1}) is with the use an auxiliary field, $\omega$, in the following form
\begin{equation}
S = \int d^4x\left( { - \frac{1}{4}\omega ^2  +
\frac{1}{2}\omega \sqrt { - F_{\mu \nu }^{a} F^{a\mu \nu } } }
\right) \quad , \quad \omega  = \varepsilon ^{\mu \nu \alpha \beta } \partial _{\left[
\mu  \right.} A^{a}_{\left. {\nu \alpha \beta } \right]}
\label{eq:sib3}
\end{equation} 
this definition introduces a new degree of freedom, the three index potential and it generates the 4-index field strength $F^{a}_{\mu \nu \alpha \beta }  \equiv \partial _{\left[ \mu  \right.} A^{a}_{\left. {\nu \alpha \beta } \right]}$. Preforming the variation of this new action with respect to the three field yields the following equation of motion for $A^{a}_{\nu\alpha\beta}$ 
\begin{equation}
\varepsilon ^{\gamma \delta \alpha \beta } \partial _\beta  \left(
{\omega  - \sqrt { - F^{a\mu \nu } F^{a}_{\mu \nu } } } \right) = 0
\label{eq:sib7}
\end{equation}
when integrated we get
\begin{equation}
\omega  = \sqrt { - F^{a}_{\mu \nu } F^{a\mu \nu } }  + f 
\label{eq:sib8}
\end{equation}
The integration constant $f$ spontaneously breaks the scale
invariance, since both $\omega$ and $\sqrt { - F^{a\mu \nu }
F^{a}_{\mu \nu } }$ transform under scale transformation but $f$ does not\footnote{$f$ has the same dimensions as the field strength $F^{a}_{\mu\nu}$, that is, dimensions of $\left( {length}\right)^{ - 2}$}. Furthermore, the variation of (\ref{eq:sib3}) with respect to the $A_{\mu\nu}^a$ field, together with (\ref{eq:sib8}), a non-Abelian generalization of (\ref{eq:sib9}).   

To see what are the application and properties of this non-linear Maxwell term in curved space time, we will begin by coupling (\ref{eq:GG}) to ordinary Einstein-gravity. One example is shown in [\refcite {SH.HH}], where the 'square root' of the Maxwell term $\sqrt{\varepsilon F_{\mu\nu }F^{\mu \nu }}$ was coupled with gravity in order to search for possible linear potentials such the ones found in flat-space (\ref{eq:sib13}). The authors found that in the presence of magnetic charges ($\varepsilon=1$) no confining potential exists and they concluded by saying that magnetic charges acts agent confinement. On the other hand, confining field solutions are found for pure electrically charged ($\varepsilon=-1$) configuration, where one finds Nariai- Bertotti-Robinson (NBR) space-time with constant scalar curvature. In these space-times the confining potentials exist whether the standard Maxwell term, $F_{\mu\nu }F^{\mu \nu }$, is present or absent.

A $2+1-$dimensional version of Einstein gravity coupled to non-linear Maxwell theory was studied in [\refcite{SH.HH1}]. The authors examined the case of constant, circularly symmetric, electric field configuration $F_{t \theta}=E_{0}$, while the Lagrangian considered was a power law of the standard Maxwell invariant 
\begin{equation}
\left\vert F_{\mu \nu }F^{\mu \nu }\right\vert ^{k}
\end{equation}
For $k=\frac{1}{2}$ the authors give two different classes of solutions, $\Lambda =0$ and $\Lambda \neq 0$.
 The first one, $\Lambda =0$, is found to be similar to a black hole solution, with a singularity at $r=0$, namely, the curvature invariants $R$, $R_{\mu \nu }R^{\mu \nu}$ and the Kretschmann scalar are all divergent as $r$ approaches zero 
\begin{equation}
R\sim \frac{1}{r^{2}} \quad , \quad R_{\mu \nu }R^{\mu \nu}\sim \frac{1}{r^{4}} \quad , \quad R_{\mu \nu \alpha \beta}R^{\mu \nu \alpha \beta}\sim \frac{1}{r^{4}}
\end{equation} 
In addition they find the singularity at the origin, $r=0$, is also a zero for $g_{tt}$ which makes the solution a black point that confines the radial motion of a massive particles both charged and uncharged. This coincides with the solutions in $2+1$ dimensions of gravity coupled with a massless, self-interacting real scalar field \cite{4}{.} The second case, $\Lambda <0$ ($k=\frac{1}{2}$), corresponds to a
non-asymptotically flat wormhole built entirely from the cosmological
constant. 

Here we will show a ``charge hiding" behaviour opposite to the famous Misner-Wheeler ``charge without charge'' effect \cite{misner-wheeler}{,} 
in which wormholes connecting two asymptotically flat space times provide the possibility of having a non-trivial electromagnetic solutions, where the lines of force of the electric field flow from one universe to the other, without a source, giving the impression of being positively charged in one universe and negatively charged in the other universe. 
In contrast, the charge hiding
effect, will cause a genuinely charged matter source of gravity
and electromagnetism to appear electrically neutral to an external observer. In order to show this we will first find two solutions of the resulting field equations and match these solution on a light-like \cite{Ed-Al-Em-Sv}{,} and subsequently time-like branes \cite{Ed-Ma}{,} thin shells of matter and charge.  
In the case where light-like branes are present as the source of charge in the system,
sitting at the throat, there is at the throat a surface of infinite red-shift for
the coordinate time, so the wormhole is accessible only according to the weaker
definition that a ``traveller" attempting to cross from one side of the
wormhole throat to the other side can do so in a finite proper time. If we want
to assure the more stringent definition of transversability or accessibility, we
must have time-like branes. We will see, however, that these wormhole, or horned-particle, solutions lead to a violation of some energy conditions , indicating the presence of exotic matter at the throat, which can be of quantum mechanical origin.

Next we will consider the coupling of (\ref{eq:GG}) to extended theories of gravity \cite{dinamicalcopuling,Ed-Go-Ru-Ma}{,} where a non-linear function of the scalar curvature $R$ and of other higher-order invariants of the Riemann curvature tensor, such as $R_{\mu\nu}R^{\mu\nu}$, are included in the gravitational action, in addition one may also include a dilaton field, $\phi$, to implement scale invariance. These kind of theories are attracting a lot of interest as possible candidates to resolve problems in the standard cosmological models related to dark matter and dark energy, where the higher-order curvature terms usually arise in approaches to quantum gravity  and in the quantization of fields in curved space-time. Here we will consider the following extended theories
\begin{equation}
F_1(R)=R+\alpha R^2 \quad , \quad F_2(R)=R+\alpha R^2+\beta R_{\mu\nu}R^{\mu\nu}
\end{equation}
The case of a standard Maxwell theory coupled to these kind of extended theories of gravity,  
in $3+1-$dimensions, was studied in detail in [\refcite{or12a}--\refcite{or13b}] where it was found that the internal point-like singularity that typically  arises in the Reissner-Nordstr{\"o}m solution of GR is generically replaced by a region of finite non-zero area that represents the throat of a wormhole. The existence of this wormhole is an effect of the modified gravitational dynamics since the electromagnetic field does not violate any of the classical energy conditions. Therefore, these kind of theories can produce wormhole solutions without requiring exotic matter to support them. 

The final demonstration will be for the ``Two Measure Theory" (TMT).  The TMT models consider two measures of integration in the action,
the standard $\sqrt{-g}$ where $g$ is the determinant of the metric and another measure $\Phi$ independent
of the metric tensor \cite{eg1TMT1,eg1TMT2,eg1TMT3,eg1TMT4,eg1TMT5}
{.} Our focus, here, will be on the interplay between gauge field dynamics (\ref{eq:GG}) and the TMT to see whether the possibility of obtaining a confinement phase and a deconfined phase (like in the MIT bag model \cite{eg1MIT}) can be addressed in this context. 
 
When dealing with extended theories of gravity we will choose to relax the Levi-Civita condition on the connection $\Gamma$, and  obtain the field equations following the Palatini approach, in which metric and connection are regarded as two physically independent entities. This implies that both metric and connection must be determined by solving their respective equations obtained through the application of the variational principle on the action. 

\newpage
\section{Coupling to Gravity}
We will start by considering the simplest coupling to gravity of the nonlinear gauge field system
with a square root of the Maxwell term (\ref{eq:GG}). The relevant action, which can also contain a cosmological constant, is given by the following expression
\begin{equation}
\begin{alignedat}{2}
&S = \int d^4 x \sqrt{-g} \left[ \frac{R(g) - 2\Lambda}{16\pi} + L(F^2)\right]
\quad ,\quad
L(F^2) = - \frac{1}{4} F^2 - \frac{f}{2} \sqrt{\varepsilon F^2}
\label{eq:gravity+GG} \\
&F^2 \equiv F_{\alpha\beta} F_{\mu\nu} g^{\alpha\mu} g^{\beta\nu} \quad ,\quad 
F_{\mu\nu} = \partial_\mu A_\nu - \partial_\nu A_\mu
\end{alignedat}
\end{equation}
Here $R(g)$ is the scalar curvature of the space-time metric
$g_{\mu\nu}$ and $g \equiv \det\Vert g_{\mu\nu}\Vert$; the sign factor $\varepsilon = \pm 1$
in the square-root term in (\ref{eq:gravity+GG}) corresponds to ``magnetic'' or ``electric'' dominance; $f$ is a positive coupling constant. The corresponding equations of motion read:
\begin{equation}
R_{\mu\nu} - \dfrac{1}{2} g_{\mu\nu} R + \Lambda g_{\mu\nu} = 8\pi T^{(F)}_{\mu\nu} 
\label{eq:einstein-eqs}
\end{equation}
\begin{equation}
T^{(F)}_{\mu\nu} =\left( 1 + 
\varepsilon \frac{f}{\sqrt{\varepsilon F^2}}\right) F_{\mu\alpha} F_{\nu\beta} g^{\alpha\beta}
- \frac{1}{4} \left( F^2 + 2f\sqrt{\varepsilon F^2}\right) g_{\mu\nu}
\label{eq:stress-tensor-F}
\end{equation}
and
\begin{equation}
\partial_\nu \left(\sqrt{-g}\left( 1 +
\varepsilon\frac{f}{\sqrt{\varepsilon F^2}}\right) F_{\alpha\beta} g^{\mu\alpha} g^{\nu\beta}\right)=0
\label{eq:GG-eqs}
\end{equation}
Next we will look for two kinds of spherically symmetric solutions of this set of equations. We will be interested in the "electric" sector of the theory, so $\varepsilon = -1$ henceforth. It is of course possible to allows magnetic dominated configurations by choosing $\varepsilon = 1$ (as we will see in section (\ref{Magnetic Dominance})), or by taking absolute value inside the square root, but this will be harder to motivate, for example from spontaneous symmetry breaking of scale invariance, furthermore we have already mentioned that, when considering only ``square-root" term in the action, purely magnetic configurations will act against confinement.      


\subsection{Reissner-Nordstr{\"o}m-de-Sitter space-time geometry (RNdS)} 

First we look for static spherically symmetric solutions of the Reissner-Nordstr{\"o}m-de-Sitter type, in the same manner to those found for the ordinary Maxwell theory ($f=0$). When looking for static, spherically symmetric configurations we have the following forms of the metric tensor and the gauge field strength: 
\begin{equation}
ds^2 = - A(r) dt^2 + \frac{dr^2}{A(r)} + r^2 \left( d\theta^2 + \sin^2 \theta d\varphi^2\right)
\label{eq:spherical-static-0}
\end{equation} 
\begin{equation}
F_{\mu\nu} = 0  \quad \mathrm{for} \quad (\mu,\nu)\neq (0,r) \quad ,\quad
F_{0r} = F_{0r} (r) 
\label{eq:electr-static-1}
\end{equation}
And the solution of the gauge field equations of motion (\ref{eq:GG-eqs}) reads:
\begin{equation}
F_{0r} = \frac{\varepsilon_F f}{\sqrt{2}} + \frac{Q}{\sqrt{4\pi}\, r^2} 
\quad ,\quad \varepsilon_F \equiv \mathrm{sign}(F_{0r}) = \mathrm{sign}(Q) \; ,
\label{eq:cornell-sol}
\end{equation}
Namely, the electric field contain a radial
constant piece 
alongside the Coulomb term. So the property of a linear ``confining-like" potential (\ref{eq:sib13}), which was found in  flat space-time, is also present when the system (\ref{eq:GG}) is coupled to gravity. 

For the metric part, it is known that for static spherically symmetric
metrics with the associated energy-momentum tensor obeying the condition $T^0_0 = T^r_r$, which is fulfilled in the present case
(\ref{eq:electr-static-1}), it is sufficient to solve Einstein equations:
\begin{equation}
R^0_0 =-\frac{1}{2r^2} \partial_r \left( r^2 \partial_r A\right)=8\pi (T^0_0 - \dfrac{1}{2} T^\alpha_\alpha)
\label{eq:Ed-Rab}
\end{equation}
\begin{equation}
R^\theta_\theta= - \frac{1}{r^2} (A-1) - \frac{1}{r}\partial_r A=- 8\pi T^0_0
\label{eq:Ed-Rab1}
\end{equation}
In the case under consideration (\ref{eq:cornell-sol}) the solution of (\ref{eq:Ed-Rab}) and (\ref{eq:Ed-Rab1}) yields:
\begin{equation}
A(r) = 1 - \sqrt{8\pi}|Q|f - \frac{2m}{r} + \frac{Q^2}{r^2} - \frac{\Lambda_{eff}}{3} r^2 \quad , \quad \Lambda_{eff}=2\pi f^2+\Lambda
\label{eq:RN-dS+const-electr}
\end{equation}
This is a Reissner-Nordstr{\"o}m-de-Sitter-type spacetime with a \emph{dynamically generated} effective cosmological constant $2\pi f^2$. In the presence of the bare cosmological constant term in (\ref{eq:gravity+GG}) the only effect is shifting of the effective cosmological constant. The appearance in (\ref{eq:RN-dS+const-electr}) of a ``leading'' constant term different from $1$ resembles the effect on gravity produced by a spherically symmetric ``hedgehog'' configuration of a non-linear sigma-model scalar field with $SO(3)$ symmetry, that is the field of a global monopole \cite{BV-hedge,BV-hedge1}{.}


\subsection{Generalized Levi-Civita-Bertotti-Robinson Space-Times (LCBR)}
Next we will look for static solutions of Levi-Civita-Bertotti-Robinson type 
\cite{LC-BR-1,LC-BR-2,LC-BR-3} of the system (\ref{eq:einstein-eqs})--(\ref{eq:GG-eqs}), namely, 
with space-time geometry of the form $\mathcal{M}_2 \times S^2$, where $\mathcal{M}_2$ is some two-dimensional manifold:
\begin{equation}
ds^2 = - A(\eta) dt^2 + \frac{d\eta^2}{A(\eta)} 
+ a^2 \left( d\theta^2 + \sin^2 \theta d\varphi^2\right)
\label{eq:gen-BR-metric}
\end{equation}
where
\begin{equation}
-\infty < \eta <\infty \quad , \quad a = \mathrm{const}
\nonumber
\end{equation}
In what follows we will find solutions for both purely electric and purely magnetic configurations of the non-linear gauge field system. For both cases the metric will have the form (\ref{eq:gen-BR-metric}), the difference is manifested in the form of the field strength tensor $F_{\mu\nu}$ and the \textit{Sign} of $\varepsilon$ in (\ref{eq:gravity+GG}).
\subsubsection{Electric Dominance}
For purely electric dominated configurations , where $\varepsilon = -1$ in (\ref{eq:gravity+GG}), the field strength tensor is of the form:
\begin{equation}
F_{\mu\nu} = 0 \quad \mathrm{for} \quad \mu,\nu\neq 0,\eta \quad ,\quad
F_{0\eta} = F_{0\eta} (\eta) 
\label{eq:electr-static}
\end{equation}
And the gauge field equations of motion yielding a globally constant vacuum electric field:
\begin{equation}
F_{0\eta} = c_F = \mathrm{arbitrary ~const}
\label{eq:const-electr}
\end{equation}
The (mixed) components of energy-momentum tensor (\ref{eq:stress-tensor-F}) read:
\begin{equation}
{T^{(F)}}^0_0 = {T^{(F)}}^\eta_\eta = - \dfrac{1}{2} F^2_{0\eta} \quad ,\quad
T^{(F)}_{ij} = g_{ij}\left(\dfrac{1}{2} F^2_{0\eta} - \frac{f}{\sqrt{2}}|F_{0\eta}|\right)
\label{eq:T-F-electr}
\end{equation}
Taking into account (\ref{eq:T-F-electr}) together with the $(ij)$ component of (\ref{eq:einstein-eqs}), where $R_{ij}=\frac{1}{a^2} g_{ij}$ because of the $S^2$ factor in
(\ref{eq:gen-BR-metric}), yield:
\begin{equation}
\frac{1}{a^2} = 4\pi c_F^2 + \Lambda 
\label{eq:einstein-ij}
\end{equation}
While the $(00)$ component of (\ref{eq:einstein-eqs}), using the
expression $R^0_0 = - \dfrac{1}{2} \partial_\eta^2 A$, becomes:
\begin{equation}
\partial_\eta^2 A = 8\pi h(|c_F|) \quad ,\quad
h(|c_F|) \equiv c_F^2 - \sqrt{2}f|c_F| - \frac{\Lambda}{4\pi} \; ,
\label{eq:einstein-00}
\end{equation}
We distinguish between 
three distinct types of Levi-Civita-Bertotti-Robinson solutions for gravity coupled to the non-linear gauge field system (\ref{eq:gravity+GG}):
\begin{itemize}
\item[(1)] $AdS_2 \times S^2$, where $AdS_2$ is two-dimensional anti-de Sitter 
space with 
\begin{equation}
A(\eta) = 4\pi h(|c_F|)\,\eta^2 \quad ,\quad h(|c_F|) >0 
\label{eq:AdS2}
\end{equation}
$\eta$ being the Poincare patch
space-like coordinate, 
provided:
\begin{equation}
|c_F| > \frac{f}{\sqrt{2}}\Bigl( 1 + \sqrt{1 + \frac{\Lambda}{2\pi f^2}}\Bigr) 
\quad \mathrm{for} \quad \Lambda \geq - 2\pi f^2
\label{eq:AdS2-cF-1}
\end{equation}

\begin{equation}
|c_F| > \sqrt{\frac{1}{4\pi}|\Lambda|} 
\quad \mathrm{for}\;\; \Lambda < 0 \;\; ,\;\; |\Lambda| > 2\pi f^2 \; .
\label{eq:AdS2-cF-3}
\end{equation}

\item[(2)] $Rind_2 \times S^2$, where $Rind_2$ is the flat two-dimensional 
Rindler space with:
\begin{equation}
A(\eta) = \eta \;\; \mathrm{for}\; 0 < \eta < \infty \quad \mathrm{or} \quad
A(\eta) = - \eta \;\; \mathrm{for}\; -\infty <\eta < 0 
\label{eq:Rindler2}
\end{equation}
provided:
\begin{equation}
|c_F| = \frac{f}{\sqrt{2}}\Bigl( 1 + \sqrt{1 + \frac{\Lambda}{2\pi f^2}}\Bigr) 
\quad \mathrm{for}\;\; \Lambda \geq - 2\pi f^2 \; ,
\label{eq:Rindler2-cF-1}
\end{equation}
\item[(3)] $dS_2 \times S^2$, where $dS_2$ is two-dimensional de-Sitter space with:
\begin{equation}
A(\eta) = 1 - K(|c_F|)\,\eta^2 \;\; ,\;\; 
K (|c_F|) \equiv 4\pi\Bigl(\sqrt{2}f |c_F| - c_F^2 +
\frac{\Lambda}{4\pi}\Bigr) >0
\label{eq:dS2}
\end{equation}
provided:
\begin{equation}
|c_F| < \frac{f}{\sqrt{2}}\Bigl( 1 + \sqrt{1 + \frac{\Lambda}{2\pi f^2}}\Bigr)
\quad \mathrm{for}\;\; \Lambda > - 2\pi f^2 \; .
\label{eq:dS2-cF-1}
\end{equation}
Note that $dS_2$ has {\em two horizons} at
\begin{equation}
\eta = \pm \eta_0 \equiv 
\pm \left[ 4\pi\left(\sqrt{2}f|c_F| - c_F^2\right) + \Lambda\right]^{-\dfrac{1}{2}}
\end{equation} 
When $\Lambda = 0$, for the special value $|c_F| = \frac{f}{\sqrt{2}}$ we recover the 
Nariai solution \cite{nariai-1,nariai-2} with $A(\eta) = 1 - 2\pi f^2 \eta^2$ 
\end{itemize}
In all three cases above there is a constant vacuum electric field $|F_{0\eta}| = |c_F|$, and the size of the $S^2$ factor is given by (\ref{eq:einstein-ij}).
Solutions (\ref{eq:Rindler2}) and (\ref{eq:dS2}) with $\Lambda=0$ are new ones and are specifically due to the presence of the, non-linear, Maxwell square-root term in the gauge field Lagrangian (\ref{eq:gravity+GG}).
\subsubsection{Magnetic Dominance}\label{Magnetic Dominance}

For purely magnetic dominated configurations, where $\varepsilon = +1$ in (\ref{eq:gravity+GG}), the field strength tensor is of the form:
\begin{equation}
F_{\mu\nu} = 0 \;\; \mathrm{for}\; \mu,\nu\neq i,j\equiv \theta,\varphi\quad ,\quad
\partial_0 F_{ij} = \partial_\varphi F_{ij} = 0
\label{eq:magnet-static}
\end{equation}
The gauge field equations of
motion (\ref{eq:GG-eqs}) yield magnetic monopole solution:
\begin{equation}
F_{ij} = B a^2\sin\theta\,\varepsilon_{ij} \;\; ,\;\; B=\mathrm{const} 
\label{eq:monopole}
\end{equation}
irrespective of the presence of the ``square-root'' Maxwell term. The latter, 
however, does contribute to the energy-momentum tensor:
\begin{equation}
{T^{(F)}}^0_0 = {T^{(F)}}^\eta_\eta = - \dfrac{1}{2} B^2 - f|B| \quad ,\quad
T^{(F)}_{ij} = \dfrac{1}{2} g_{ij} B^2 \; .
\label{eq:T-F-magnet}
\end{equation}
Taking into account (\ref{eq:T-F-magnet}), the $(ij)$ component of (\ref{eq:einstein-eqs}) yield ( in the same manner as (\ref{eq:einstein-ij})):
\begin{equation}
\frac{1}{a^2} = 4\pi\left( B^2 + \sqrt{2}f|B|\right) + \Lambda
\label{eq:einstein-ij-1}
\end{equation}
which determines the size of the $S^2$ factor.
 Here, unlike the previous case, the size of the $S^2$-factor depends on the ``square-root'' Maxwell coupling constant $f$. While 
the mixed-component $(00)$ of (\ref{eq:einstein-eqs}) gives:
\begin{equation}
\partial^2_\eta A = 8\pi B^2 - 2\Lambda 
\label{eq:einstein-00-1}
\end{equation}
Thus, in the purely magnetic case we recover the three types of
Levi-Civita-Bertotti-Robinson solutions, which where found for the electric case, only here we will have constant-magnitude magnetic field.




\subsection{General vacuum solutions}\label{General vacuum solutions}
As a final remark, returning to the non-linear gauge field Equation (\ref{eq:GG-eqs}) we see that there exists a more general \emph{vacuum} solution of the latter \emph{without} the assumption of staticity and spherical symmetry:
\begin{equation}
F^2 \equiv F_{\alpha\beta} F_{\mu\nu} g^{\alpha\mu} g^{\beta\nu} = - f^2 = \mathrm{const} 
\label{eq:const-F2}
\end{equation}
The latter automatically produces via (\ref{eq:stress-tensor-F}) an effective positive
cosmological constant:
\begin{equation}
T^{(F)}_{\mu\nu} = - \frac{f^2}{4} g_{\mu\nu} \quad ,\;\; \mathrm{i.e.} \;\;
\Lambda_{\mathrm{eff}} = 2\pi f^2
\label{eq:stress-tensor-F-vac}
\end{equation}
This reduces the gravity/gauge-field equations of motion (\ref{eq:einstein-eqs})-(\ref{eq:GG-eqs}) to the vacuum Einstein equations with effective cosmological constant:
\begin{equation}
R_{\mu\nu} - \dfrac{1}{2} g_{\mu\nu} R + (\Lambda + 2\pi f^2) g_{\mu\nu} = 0
\label{eq:einstein-eqs-vac}
\end{equation}
supplemented with the constraint (\ref{eq:const-F2}).
Thus, assuming absence of magnetic field ($F_{mn}=0$), \textsl{i.e.}, 
$F^2 = 2 E_m E_n G^{mn} G^{00}\; ,\; E_m \equiv F_{0m}\; (m,n=1,2,3)$, we obtain an electrically neutral Schwarzschild-(anti)-de-Sitter or purely
Schwarzschild solutions with a constant vacuum electric field, 
which according to (\ref{eq:const-F2}) has constant magnitude:
\begin{equation}
|\vec{E}| \equiv \sqrt{-\dfrac{1}{2} F^2} = \frac{f}{\sqrt{2}}
\label{eq:const-magnitude}
\end{equation}
but it may point in arbitrary direction. In this vacuum with disordered 
constant-magnitude 
electric field it will not be able to pass energy to a test charged particle,
which instead will undergo a kind of Brownian motion, therefore {\em no} Schwinger
pair-creation mechanism will take place.


\newpage
\section{Lightlike Brane Sources}\label{Lightlike Brane Sources}
In this section we will match the two solutions we have found on a spherically symmetric light-like shell. \textsl{LL-brane} (also called null-branes) are of substantial interest
in general relativity as they describe impulsive lightlike signals arising
in various violent astrophysical events, e.g., final explosion in cataclysmic processes
such as supernovae and collision of neutron stars. \textsl{LL-brane} also play
important role in the description of various other physically important cosmological
and astrophysical phenomena such as the ``membrane paradigm" of black
hole physics and the thin-wall approach to domain walls coupled to gravity. Before proceeding with the matching of the two solutions, (\ref{eq:electr-static-1}) and (\ref{eq:gen-BR-metric}), we will first present the \textsl{LL-brane} action and the conditions for ``sewing" two solutions on it. 
\subsection{\textsl{LL-brane} action}
World-volume \textsl{LL-brane} actions in a reparametrization-invariant Nambu-Goto-type or in an equivalent Polyakov-type formulation were proposed in [\refcite{KerrWH-varna2008-rotWH-ERbridge-BRkink}], and are given by:
\begin{equation}
S_{\rm LL}\left[ q\right] = - \dfrac{1}{2} \int d^{p+1}\sigma T b_0^{\frac{p-1}{2}}\sqrt{-\gamma}\left[ \gamma^{ab} {\bar g}_{ab} - b_0 (p-1)\right]
\label{eq:LL-action+EM}
\end{equation}
\begin{equation}
{\bar g}_{ab} \equiv \partial_a X^\mu g_{\mu\nu} \partial_b X^\nu 
- \frac{1}{T^2} (\partial_a u + q\mathcal{A}_a)(\partial_b u  + q\mathcal{A}_b) 
\quad , \quad \mathcal{A}_a \equiv \partial_a X^\mu A_\mu 
\label{eq:ind-metric-ext-A}
\end{equation}
Here and below the following notations are used:
\begin{itemize}
\item
$\gamma_{ab}$ is the {\em intrinsic} world-volume Riemannian metric.
\item
$g_{ab}=\partial_a X^{\mu} g_{\mu\nu}(X) \partial_b X^{\nu}$ is the {\em induced} metric on the 
world-volume, which becomes {\em singular} on-shell (manifestation of the lightlike 
nature).
\item
$b_0$ is world-volume ``cosmological constant''.
\item
$X^\mu (\sigma)$ are the $p$-brane embedding coordinates in the bulk
$D$-dimensional spacetime with Riemannian metric
$g_{\mu\nu}(x)$ ($\mu,\nu = 0,1,\ldots ,D-1$); 
$(\sigma)\equiv \left(\sigma^0 \equiv \tau,\sigma^i\right)$ with $i=1,\ldots ,p$ ;
$\partial_a \equiv \dfrac{\partial}{\partial\sigma^a}$.
\item
$u$ is auxiliary world-volume scalar field defining the lightlike direction
of the induced metric;
\item
$T$ is {\em dynamical (variable)} brane tension;
\item
$q$ -- the coupling to bulk spacetime gauge field $\mathcal{A}_\mu$ is
\textsl{LL-brane} surface charge density.
\end{itemize}
The on-shell singularity of the induced metric $g_{ab}$ , \textsl{i.e.}, 
the lightlike property, directly follows from the \textsl{LL-brane} equations of
motion:
\begin{equation}
g_{ab}\left( {\bar g}^{bc}\left(\partial_c u  + q\mathcal{A}_c\right)\right) = 0 
\label{eq:on-shell-singular-A}
\end{equation}


\subsection{Field Equations And Matching Conditions}
Now, the full action of the self-consistently coupled gravity, non-linear gauge field and \textsl{LL-brane} system reads;
\begin{equation}
S = \int d^4 x \sqrt{-g} \left[\frac{R(G) - 2\Lambda}{16\pi} + L(F^2)\right]
+ \sum_{k=1}^N S_{\mathrm{LL}}[q^{(k)}]
\label{eq:gravity+GG+LL}
\end{equation}
where the superscript $(k)$ indicates the $k$-th \textsl{LL-brane}, and $L(F^2)$ is given by (\ref{eq:GG}). 
The corresponding equations of motion are as follows:
\begin{equation}
R_{\mu\nu} - \dfrac{1}{2} g_{\mu\nu} R + \Lambda g_{\mu\nu} = 
8\pi \left[ T^{(F)}_{\mu\nu} + \sum_{k=1}^N T^{(k)}_{\mu\nu}\right]
\label{eq:einstein+LL-eqs}
\end{equation}
\begin{equation}
\partial_\nu \left[\sqrt{-g} \left( 1 - \frac{f}{\sqrt{-F^2}}\right) 
F_{\alpha\beta} g^{\mu\alpha} g^{\nu\beta}\right] + \sum_{k=1}^N j_{(k)}^\mu = 0
\label{eq:GG+LL-eqs}
\end{equation}
The energy-momentum tensor and the charge current density of $k$-th 
\textsl{LL-brane} are straightforwardly derived from the \textsl{LL-brane} 
world-volume action (\ref{eq:LL-action+EM}) and are given by:
\begin{equation}
T_{(k)}^{\mu\nu} = 
- \int d^3\sigma\dfrac{\delta^{(4)}\left( x-X_{(k)}(\sigma)\right)}{\sqrt{-g}}
T^{(k)}\sqrt{|{\bar g}_{(k)}|} {\bar g}_{(k)}^{ab}
\partial_a X_{(k)}^\mu \partial_b X_{(k)}^\nu 
\label{eq:T-brane-A}
\end{equation}
\begin{equation}
j_{(k)}^\mu = -
q^{(k)} \int d^3\sigma\delta^{(4)}\left( x-X_{(k)}(\sigma)\right)
\sqrt{|{\bar g}_{(k)}|} {\bar g}_{(k)}^{ab}\partial_a X_{(k)}^\mu 
\frac{\partial_b u^{(k)} + q^{(k)}\mathcal{A}^{(k)}_b}{T^{(k)}}
\label{eq:j-brane-A}
\end{equation}
Solving (\ref{eq:einstein+LL-eqs})--(\ref{eq:GG+LL-eqs}) with (\ref{eq:T-brane-A})--(\ref{eq:j-brane-A})
one finds ``thin-shell'' wormhole solutions of static ``spherically-symmetric'' type of the form  
(in Eddington-Finkelstein coordinates 
$dt=dv-\dfrac{d\eta}{A(\eta)}\; ,\; F_{0\eta} = F_{v\eta}$):
\begin{equation}
ds^2 = - A(\eta) dv^2 + 2dv d\eta + C(\eta) h_{ij}(\theta) d\theta^i d\theta^j \quad ,\quad F_{v\eta} = F_{v\eta} (\eta)
\label{eq:static-spherical-EF}
\end{equation}
\begin{equation}
-\infty < \eta < \infty \quad, \;\; A(\eta^{(k)}_0) = 0 \;\; 
\mathrm{for} \;\; \eta^{(1)}_0 <\ldots<\eta^{(N)}_0 
\label{eq:common-horizons}
\end{equation}
Where the derivation of these ``thin-shell'' wormhole solutions proceeds along the
following main steps:

\begin{itemize}
\item[(i)] Take ``vacuum'' solutions of (\ref{eq:einstein+LL-eqs})--(\ref{eq:GG+LL-eqs})
(without delta-function \textsl{LL-brane} terms) in each spacetime region
(separate ``universe'') given by 
$\bigl(-\infty\! <\!\eta\!<\!\eta^{(1)}_0\bigr),\ldots,$
$\bigl(\eta^{(N)}_0 \!<\!\eta \!<\!\infty\bigr)$ with common horizon(s) at 
$\eta=\eta^{(k)}_0$ ($k=1,\ldots ,N$).
\item[(ii)] Each $k$-th \textsl{LL-brane} automatically locates itself on the
horizon at $\eta=\eta^{(k)}_0$, this is an intrinsic property of \textsl{LL-brane} dynamics.
\item[(iii)] Match discontinuities of the derivatives of the metric and
the gauge field strength  across each horizon at $\eta=\eta^{(k)}_0$ using the 
explicit expressions for the \textsl{LL-brane} 
stress-energy tensor and charge current density (\ref{eq:T-brane-A})--(\ref{eq:j-brane-A}).
\end{itemize}


\subsection{Charge ``Hiding'' as a consequence of a non-trivial topology}
First we will consider the matching of a non-compact ``universe'', RNdS geometry (\ref{eq:RN-dS+const-electr}), to a compactified,``tube-like'', ``universe'' of LCBR type with geometry $AdS_2 \times S^2$ or $Rind_2 \times S^2$ ((\ref{eq:AdS2}),  (\ref{eq:Rindler2}) respectively), with the charged \textsl{LL-brane} occupying the ``throat''.  This configuration, which has wormhole-like geometry, is described as "horned-particle" as done for the space-times studied in [\refcite{horned-particle-1,horned-particle-2}] which have large "tube" like structure also. Somewhat similar are the so called "Gravitational bags", where some extra dimensions grows very large at the center of the four dimensional projected metric \cite{GB-123,GB-124,GB-125}{.} The horned-particle geometry, which makes the topological hiding effect possible, is visualized on Fig.1 below.
\begin{figure}[H]
\begin{center}
\includegraphics[scale=0.8,angle=270,keepaspectratio=true]{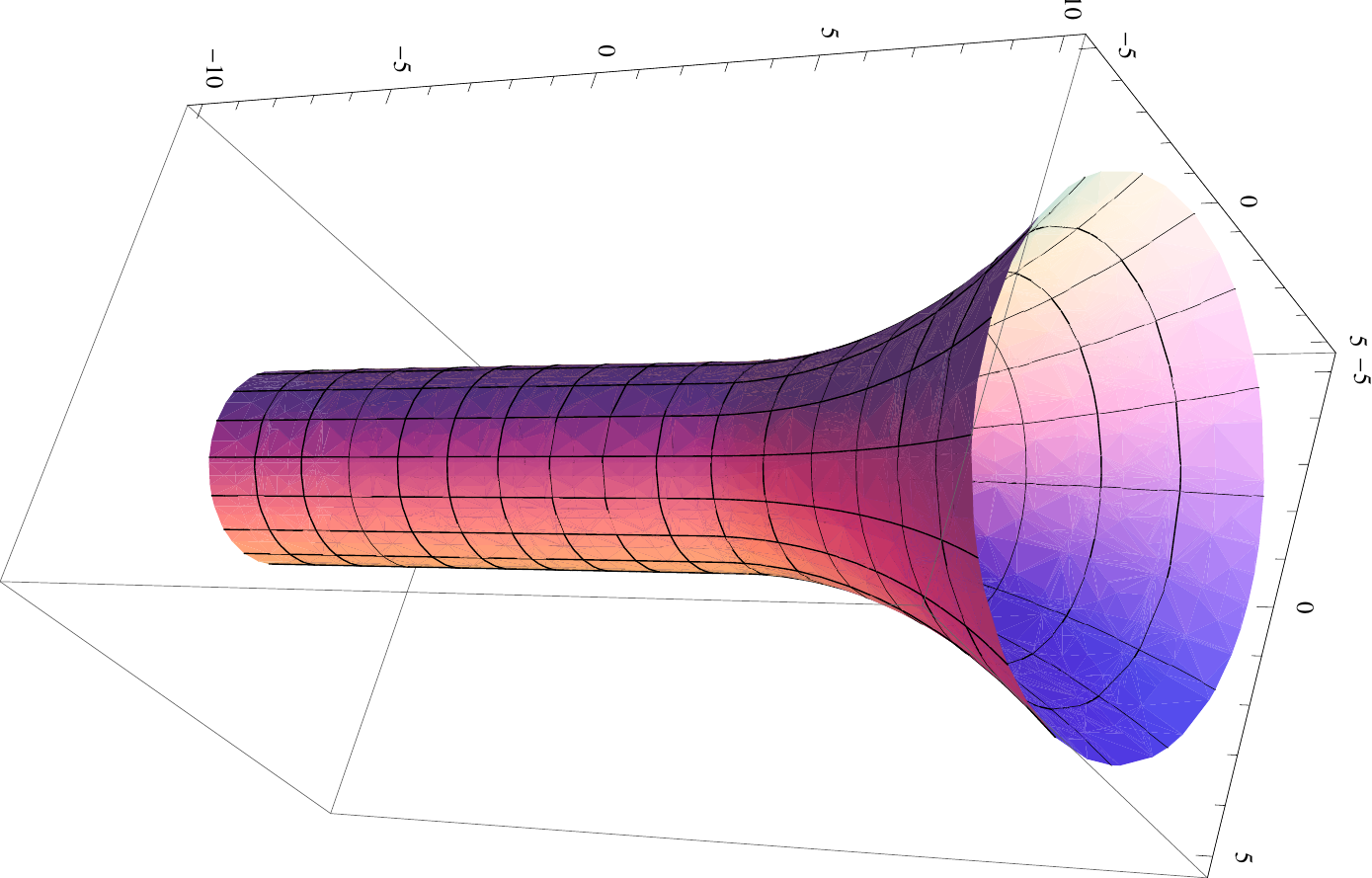}
\caption{Shape of $t=const$ and $\theta=\frac{\pi}{2}$ slice of
{\em charge-``hiding''}, horned-particle geometry: the whole electric flux produced
by the charged \textsl{LL-brane} at the ``throat'' is expelled 
into the left infinitely long cylindric tube.}
\label{fig:HornedParticle}
\end{center}
\end{figure}
The matching relations for the discontinuities of the metric and gauge field
components across the \textsl{LL-brane} world-volume occupying the
wormhole ``throat'' (which are here derived self-consistently from a
well-defined world-volume Lagrangian action principle for the \textsl{LL-brane}
(\ref{eq:LL-action+EM})) determine all parameters of the horned-particle solutions as
functions of $q$ (the \textsl{LL-brane} charge) and $f$ (coupling constant of 
$\sqrt{-F^2}$), this turn out to be:
\begin{equation}
Q=0 \quad ,\quad |c_F| = |q| + \frac{f}{\sqrt{2}} \quad , \quad -4\pi\left(|q|+\frac{f}{\sqrt{2}}\right)^2 < 
\Lambda_0 < 4\pi\left(q^2 -\frac{f^2}{2}\right) 
\label{eq:param-1}
\end{equation}
Since $Q=0$, this horned-particle configuration possess the novel property of {\em hiding} electric charge from external observer in the non-compact ``universe''. In other words, the whole electric flux ,produced by the charged \textsl{LL-brane} at the horned-particle ``throat'', is pushed into the ``tubelike'' ``universe'', therefore, an external observer in the non-compact ``universe'' detects a {\em genuinely charged} matter source (the charged \textsl{LL-brane}) as {\em electrically neutral}. As a result, the non-compact ``universe'' becomes electrically neutral with Schwarzschild-(anti-)de-Sitter or purely Schwarzschild geometry  carrying a vacuum constant radial electric field with magnitude given in (\ref{eq:const-magnitude}).


\subsection{Charge ``Confining" as a consequence of a non-trivial topology}
Further, we find  more interesting ``two-throat'' horned-particle solution exhibiting 
{\em QCD-like charge confinement} effect -- obtained from 
two identical {\em oppositely charged} \textsl{LL-branes} ((\ref{eq:gravity+GG+LL}) with $N=2$). The total ``two-throat''
horned-particle space-time manifold is made of (\ref{eq:RN-dS+const-electr}) and (\ref{eq:dS2}). As we have seen (\ref{eq:dS2}) has two horizons, therefore one \textsl{LL-brane} will automatically occupy the first horizon while the second \textsl{LL-brane} automatically occupies the other horizon. A simple visualization of this {\em ``two-throat'' } horned-particle geometry is given in Fig.2.
\begin{figure}[H]
\begin{center}
\includegraphics[scale=0.8,angle=270,keepaspectratio=true]{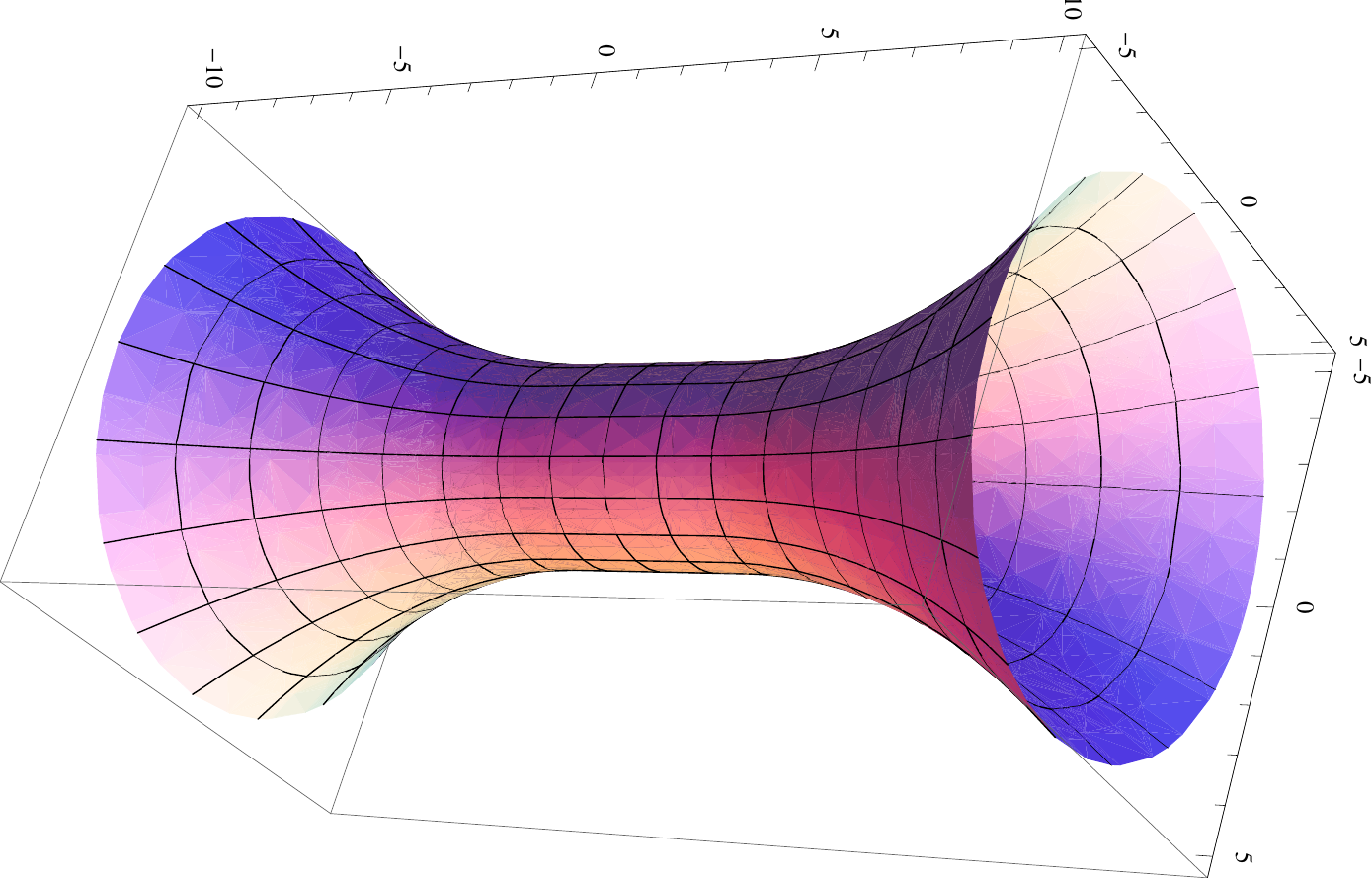}
\caption{Shape of $t=const$ and $\theta=\frac{\pi}{2}$ slice of
{\em charge-confining}, ``two-throat'' horned-particle geometry: the whole electric flux produced
by the two oppositely charged \textsl{LL-branes} is confined within the middle finite-extent cylindric tube between the ``throats''.}
\end{center}
\end{figure}
Again, the matching relations for the discontinuities of the metric and gauge field
components across each of the two \textsl{LL-brane} world-volumes
determine all the parameters of the solutions as functions of $\pm q$
( since the \textsl{LL-branes} have opposite charges) and $f$ (coupling constant of 
$\sqrt{-F^2}$), this is found to be (for each \textsl{LL-brane}):
\begin{equation}
Q=0 \quad ,\quad |c_F| = |q| + \frac{f}{\sqrt{2}} \quad , \quad 
\label{eq:param-1-conf}
\end{equation}
and the bare cosmological constant must be in the interval:
\begin{equation}
\Lambda \leq 0 \quad ,\quad |\Lambda| < 2\pi (f^2 - 2 q^2) 
\quad \to \quad |q| < \frac{f}{\sqrt{2}}
\label{eq:CC-interval-conf}
\end{equation}
Therefore, just like in the previous case, we conclude that the ``left-most'' and ``right-most'' non-compact ``universes'' become
two identical copies of the {\em electrically neutral} exterior region of
Schwarzschild-de-Sitter black hole beyond the Schwarzschild horizon. They both
carry a constant vacuum radial electric field with magnitude (\ref{eq:const-magnitude}). The corresponding electric displacement field, in the ``left-most" and ``right-most" regions, given by 
\begin{equation}
\vec{D} = \left( 1 - \frac{f}{\sqrt{2}|\vec{E}|}\right)\vec{E}
\end{equation}
is zero, so there is {\em no} electric flux in those regions, the whole electric flux produced by the two charged \textsl{LL-branes}, with 
opposite charges $\pm q$, at the boundaries of the above non-compact ``universes''
is {\em confined} within the middle, ``tube-like'', ``universe'' of LCRB type with geometry $dS_2 \times S^2$ (\ref{eq:dS2}). The ``tube-like'' region, thus, has a constant electric field 
\begin{equation}
|\vec{E}|=\frac{f}{\sqrt{2}} + |q|
\end{equation}
which is associated with a non-zero electric displacement field $|\vec{D}|= |q|$ . This is a topologically induced {\em QCD-like confinement}.


\subsection{Properties}
These horned-particle solutions share the following important properties: (i) The horned-particle space-times constructed via \textsl{LL-branes} at
their ``throats'' are {\em not} traversable w.r.t. the ``laboratory'' time of a 
static observer in either of the different ``universes'' comprising the pertinent 
 horned-particle space-time manifold since the \textsl{LL-branes} sitting at the
``throats'' look as black hole horizons to the static observer. Whoever, 
they {\em are traversable} w.r.t. the {\em proper time} of a travelling (co-moving) observer. (ii) The \textsl{LL-branes} at the  horned-particle ``throats'' represent ``exotic'' 
matter with negative or zero brane tension implying violation of null-energy conditions as predicted by general wormhole arguments \cite{visser-book} (although the latter could be remedied via quantum fluctuations).

\newpage
\section{Timelike Brane Sources}\label{Timelike Brane Sources}
Now, we want to discuss the matching of (\ref{eq:electr-static-1}) to (\ref{eq:gen-BR-metric}) at a spherically symmetric time-like wall. The equations of motion of a thin layer in GR have been obtained by W.Israel \cite{Israel-66}{,} we will briefly review these results and apply them to our case.


\subsection{Thin shell Formalisem}

To get the matching conditions it is useful to define a Gaussian Normal Coordinate system in a neighbourhood of the wall as follows; denoting the $2+1$ dimensional hyper-surface $\Sigma$ and introducing a coordinate system on $\Sigma$, two are taken to be the angular variables $\theta,\phi$, which are always well defined up to an overall rotation for a spherically symmetric configuration. For the other coordinate in the wall, one can use the proper time variable $\tau$ that would be measured by an observer moving along with the wall. The fourth coordinate $\xi$ is taken as the proper distance along the geodesics intersecting $\Sigma$ orthogonally. We adopt the convention that $\xi$ is taken to be positive in the Reissner-Nordstr{\"o}m-de-Sitter-type regime and negative in the Generalized Levi-Civita-Bertotti-Robinson regime, and $\xi=0$ is of course the position of the wall. Thus the full set of coordinates is given by $x^\mu=(\tau,\theta,\phi,\xi)$,   in this coordinates 
\begin{equation}
g^{\xi\xi}=g_{\xi\xi}=1 \quad , \quad g^{\xi i}=g_{\xi i}=0 
\end{equation}
Also, we define $n_{\mu}$ to be the normal to an $\xi=constant$ hypersurface, which in Gaussian normal coordinates has the simple form $n_{\mu}=n^{\mu}=(0,0,0,1)$. We then define the extrinsic curvature corresponding to each $\xi=constant$ hypersurface, which is a $3$-dimensional tensor whose components are 
\begin{equation}
K_{ij}=n_{i;j}=\dfrac{\partial n_i}{\partial x^j}-\Gamma^{\alpha}_{ij}n_{\alpha}=-\Gamma^{\xi}_{ij}=\dfrac{1}{2}\partial_{\xi}g_{ij} 
\label{eq:extrinsic curvature}
\end{equation}
As we can see, the extrinsic curvature gives the change of the metric in the direction perpendicular to the surface. In terms of these variables, the Einstein's equations take the form
\begin{equation}
 \begin{array}{l l}
  G^{\xi}_{\xi}\equiv -\frac{1}{2}{^{(3)}R}+\frac{1}{2}\left[(Tr K)^2-Tr(K^2) \right]=8\pi G T^{\xi}_{\xi} & \\ \\
  G^{\xi}_{i}\equiv K^{m}_{i|m}-(TrK)_{|i}=8\pi G T^{\xi}_{i}
  \end{array} 
  \label{eq:first Einstein FQ}
\end{equation}
and 
\begin{equation}
 \begin{array}{l l}
G^{i}_{j}\equiv {^{(3)}G^{i}_{j}}-\left(K^{i}_{j}-\delta^{i}_{j} TrK\right)_{,\xi} -\left(Tr K\right)K^{i}_{j} &\\ \\ \qquad +\frac{1}{2}\delta^{i}_{j} \left[(Tr K)^2+Tr(K^2) \right]=8\pi G T^{i}_{j}
 \end{array} 
  \label{eq:second Einstein FQ}
\end{equation}
where the subscript vertical bar denotes the $3$-dimensional covariant derivative in the $2+1$ dimensional space of coordinates $(\tau,\theta,\phi)$, and comma denotes an ordinary derivative. Also quantities like $^{(3)}R$, $^{(3)}G^{i}_{j}$, etc. are to be evaluated as if they concerned to a purely $3$-dimensional metric $g_{ij}$, without any reference as to how it is embedded in the higher four dimensional space. 

By definition, for a thin wall, the energy-momentum tensor $T^{\mu\nu}$ has a delta function singularity at the wall, so one can define a surface stress energy tensor $S^{\mu\nu}$
\begin{equation}
T^{\mu\nu}=S^{\mu\nu}\delta(\xi)+regular\space\ terms
\label{eq:Energy momentum tensor}
\end{equation}
When this energy momentum tensor is inserted into the field equations (\ref{eq:first Einstein FQ}) and (\ref{eq:second Einstein FQ}) we are leads to the discontinuity condition 
\begin{equation}
S^i_j=-\frac{1}{8\pi G}\left( \gamma^i_j-\delta^i_j Tr\gamma \right)
\label{eq:TTcomponent}
\end{equation}
where
\begin{equation}
 \gamma_{ij}=\lim_{\epsilon\rightarrow0}\left[K_{ij}(\xi=+\epsilon)-K_{ij}(\xi=-\epsilon) \right]
\end{equation}
is the \emph{jump} of extrinsic curvature across $\Sigma$.

In addition to this, the timelike brane also have a delta function charge density $j^\mu=\delta ^{\mu}_0 q \delta(\xi)$ coming from the discontinuity in the gauge field strength across the matching 
\begin{equation}
[F_{0\nu}]_{\Sigma}=q
\end{equation}
which can also be defined in terms of the electric  displacement field $D_{0\nu}$ in the two regions which in the present case is significantly different from the electric field $F_{0\nu}$ due to the presence of the "square-root" Maxwell term
\begin{equation}
 D_{0r}|_{\Sigma}-D_{0\eta}=q
  \label{eq:gauge field discontinuity}
\end{equation}
where
\begin{equation}
D_{0r}=\left(1-\dfrac{f}{\sqrt{2}|F_{0r}|}\right)F_{0r} \quad , \quad D_{0\eta}=\left(1-\dfrac{f}{\sqrt{2}|F_{0\eta}|}\right)F_{0\eta}
\end{equation}
This completes the matching conditions of the two solutions on a charged timelike brane.  


\subsection{Matching The Solutions on a Time-Like Brane}
The metric induced at the wall has to be well defined, and the coefficient of purely angular displacements $d\Omega^2$ in (\ref{eq:electr-static-1}) and (\ref{eq:gen-BR-metric}) has to agree at the position of the wall, to give the same value of $ds^2$,  therefore at the wall we have
\begin{equation}
r=a
\label{eq:location}
\end{equation}
The shell has surface stress-energy content  
\begin{equation}
S_{ab}=\sigma U_aU_b+p(g_{ab}+U_aU_b)
\label{eq:surface energy tensor}
\end{equation} 
For a given equation of state, $p=p(\sigma)$, the local energy momentum conservation at the wall gives $d(r^2\sigma)=-pd(r^2)$, but since we have $r=Const$ at the matching point, we get $\sigma=Const$ and from the generic equation of state $p=p(\sigma)=Const$. So the matching conditions, for the metrics (\ref{eq:electr-static-1}) and (\ref{eq:gen-BR-metric}), are reduced to
\begin{equation}
\sqrt{A_{0}(a)}=-4\pi \sigma a
 \label{eq:theta component}
\end{equation}
where $-A_0$ denotes the $g_{00}$ component of the metric "outside", on the $r>a$ region.
So, if we are dealing with a standard static space time outside (for example in the case of Schwarzschild space using only region I), the above equation implies $\sigma<0$, that is, the matching of the tube space time with the "normal" outside space will require negative energy densities as in the charged LL-brane.

And the proper time dependence of $\eta$, coming from the $K_{\tau}^\tau$ part of (\ref{eq:TTcomponent}), is found to be 
\begin{equation}
\dot{\eta}^2=(\Delta \eta-E)^2-A_i(\eta)
\label{eq:eta potential-1}
\end{equation}
where
\begin{equation}
\Delta \equiv \frac{1}{4\pi \sigma a^4}(ma -Q^2-\Lambda_{eff}a^4)+4\pi (\sigma +2p) 
\label{eq:eta potential}
\end{equation}
and $-A_i(\eta)$ is the $g_{00}$ component of the metric ``inside", on the $r<a$ region corresponding to the compactified space (LCBR space).


\subsection{Hiding The Electric Flux}

Let us assume our ground state is just flat space, and on top of that we would like to build finite energy excitations. In the first place this means that $\Lambda_{eff}=0$ in (\ref{eq:RN-dS+const-electr}), but still, this is not enough to ensure asymptotic flatness of finite energy. Indeed, if $Q \neq 0$, the leading behaviour of the metric (\ref{eq:RN-dS+const-electr}) is not flat, but rather "hedhehog type", which have energy momentum tensor that decrease only as the square of the radius for large distances, which of course are infinite energy solutions.

This is of course consistent with the notion that in a confining theory an isolated charge  has an associated infinite energy. Here, if we add the horned particle to the problem, the isolated charge can have finite energy, provided it sends all the electric flux it produces to the tube region, this requires the vanishing of the external Coulomb part of the electric field, or $Q=0$. 
We thus find horned particle solutions (Figure \ref{fig:HornedParticle}) where all the electric flux of the charged particle at the throat flows into the tube region, furthermore, these are the only finite energy solutions. In addition the horn region of the particle is completely accessible from the outside region, containing the $r\rightarrow\infty$ space, and vice versa, according to not only the traveller that goes through the shell (as was the case for the LLB), but also to a static external observer. As in the LLB scenario, the matching requires negative surface energy density for the shell sitting at the throat.


\subsection{Stable and Unstable solutions}
From the above discussion, $\Delta$ takes the following simple form
\begin{equation}
\Delta\rightarrow \frac{m}{4\pi \sigma r_0^3}+4\pi (\sigma -2 \omega)
\end{equation}
and (\ref{eq:einstein-00}) becomes
\begin{equation}
 \partial_\eta^2 A_i(\eta)=8\pi D(|c_F|)^2  
\end{equation}
where $D(|c_F|)$ is the displacement field in the compactified region, for simplicity we will choose a solution of an anti-de-Sitter form 
\begin{equation}
A_i(\eta)=1+4\pi D^2\eta^2
\label{eq:AntiDeSitter}
\end{equation}
And (\ref{eq:eta potential-1}) can be cast onto the form 
\begin{equation}
\dot{\eta}^2+(4\pi D^2-\Delta^2)\left(\eta+\frac{\Delta E}{4\pi D^2-\Delta^2}\right)^2=\frac{4\pi D^2 E^2}{4\pi D^2-\Delta^2}-1
\end{equation}
Which has the following solutions
\begin{figure}[H]
\minipage{0.5\textwidth}
  \includegraphics[scale=0.4,keepaspectratio=true]{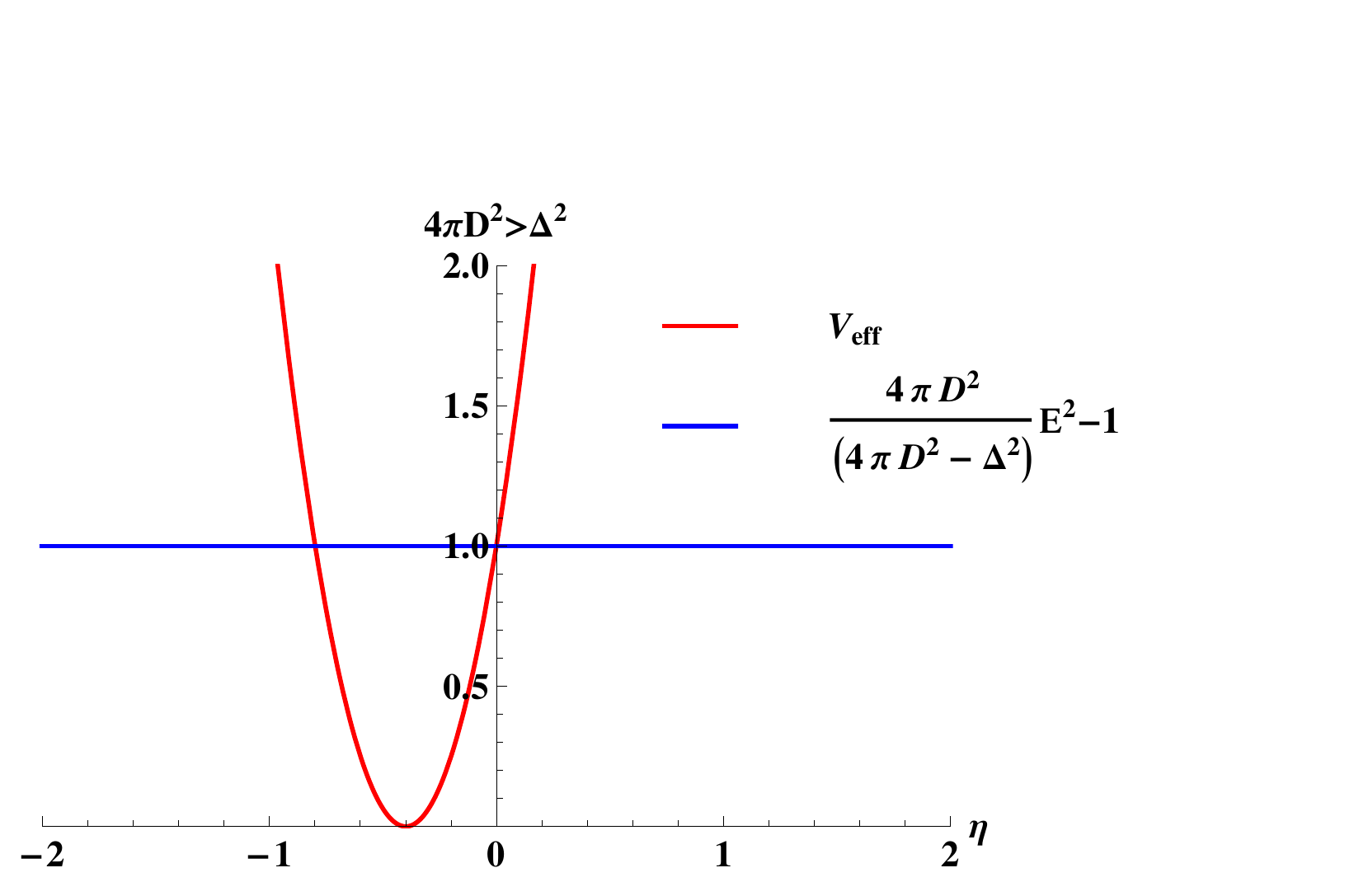}
  \caption{stable}\label{fig:stable}
\endminipage\hfill
\minipage{0.5\textwidth}
  \includegraphics[scale=0.4,keepaspectratio=true]{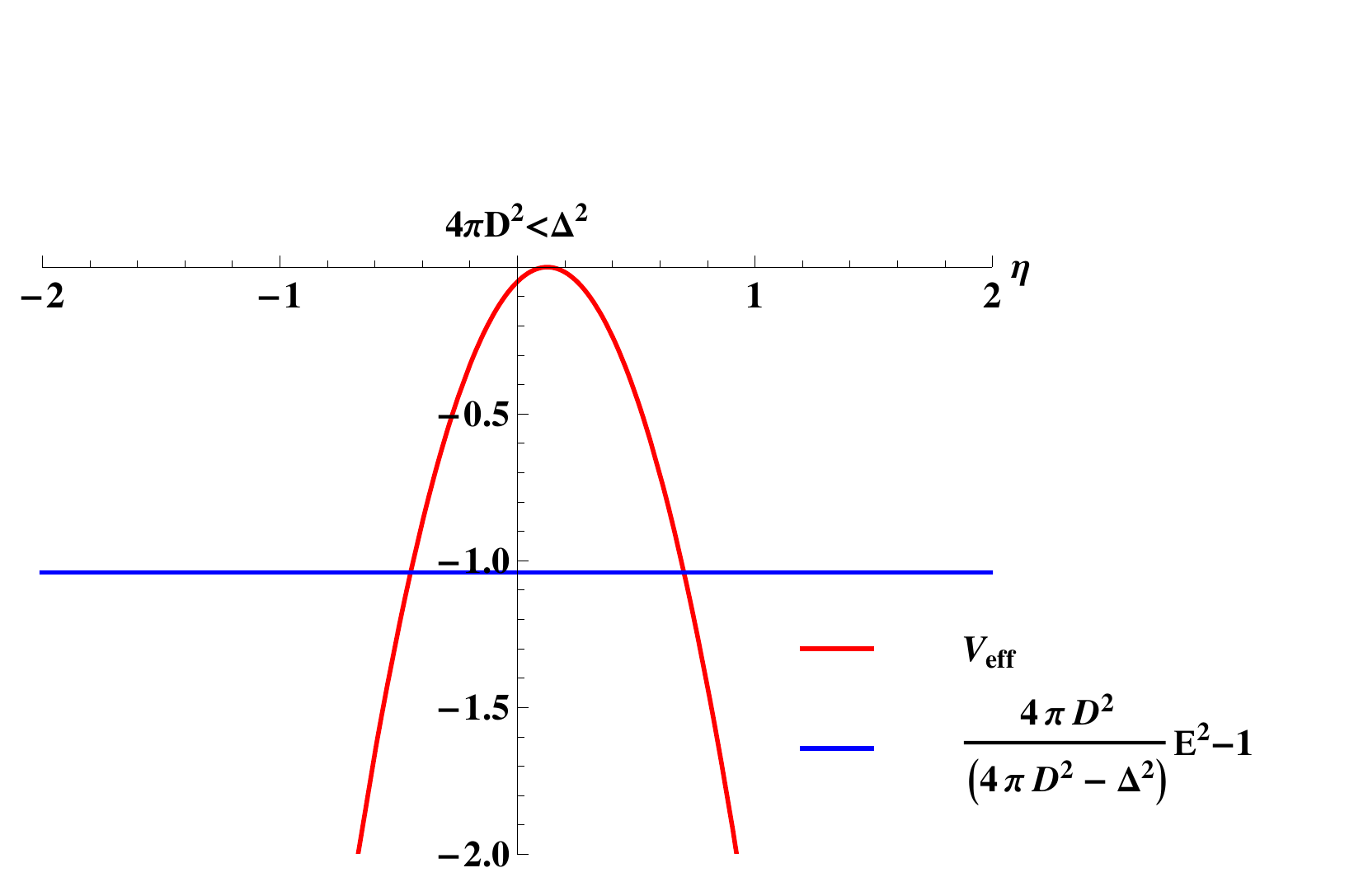}
  \caption{unstable}\label{fig:unstable}
\endminipage\hfill
\end{figure}
An illustration of the possible solutions; in the first figure (Fig. (\ref{fig:stable})), we have a stable solution with the "Energy" bounded $E^2>1-\left(\frac{\Delta}{\sqrt{4\pi}D}\right)^2\equiv E^2_{min}$. The second figure (Fig. (\ref{fig:unstable})), present an unstable solution. In this case there is no bound on the "Energy" since $\dfrac{4\pi D^2 E^2}{4\pi D^2-\Delta^2}-1<0$ for all values of $E$.  

\newpage
\section{Non-linear Maxwell term and Palatini theories}\label{Applications To Extended GR}
In this chapter we will consider static spherically symmetric electrovacuum solutions in an extension of General Relativity including $R^2$ and Ricci-squared terms, motivated by the existence of higher-order curvature corrections
in different approaches to quantum gravity, and formulated in Palatini formalism. These solutions are supported by a Maxwell electric field plus a nonlinear electromagnetic extension (\ref{eq:GG}) which here, for simplification and notation, takes the form
\begin{equation}
\varphi(X)=X -g\sqrt{2X}  \quad , \quad X=-\frac{1}{2} F_{\mu\nu}F^{\mu\nu} 
\label{eq:lagrangian-confining}
\end{equation}


\subsection{General formalism}
The action for this part is given by the following expression  
\begin{equation}
S[g,\Gamma,\psi_m]=\frac{1}{2\kappa^2}\int d^4x \sqrt{-g}f(R,Q) +\frac{1}{8\pi} \int d^4x \sqrt{-g} \varphi(X)
\label{eq:action}
\end{equation}
\begin{equation} \label{eq:gravitytheory}
f(R,Q)=R + l_P^2(a R^2+ b Q) \nonumber
\end{equation}
where $\kappa^2$ is a constant with suitable dimensions (in GR, $\kappa^2 \equiv 8\pi G$),  $\Gamma \equiv \Gamma_{\mu\nu}^{\alpha}$ is the independent connection, $g_{\alpha\beta}$ is the space-time metric,  $R=g^{\mu\nu}R_{\mu\nu}$, $Q=g^{\mu\alpha}g^{\nu\beta}R_{\mu\nu}R_{\alpha\beta}$, $R_{\mu\nu}={R^\rho}_{\mu\rho\nu}$, ${R^\alpha}_{\beta\mu\nu}=\partial_{\mu}
\Gamma^{\alpha}_{\nu\beta}-\partial_{\nu}
\Gamma^{\alpha}_{\mu\beta}+\Gamma^{\alpha}_{\mu\lambda}\Gamma^{\lambda}_{\nu\beta}-
\Gamma^{\alpha}_{\nu\lambda}\Gamma^{\lambda}_{\mu\beta} $ is the Riemann tensor, $l_P \equiv \sqrt{\hbar G/c^3}$ is the Planck length, and $a$ and $b$ are dimensionless parameters which for simplification are taken to be $b=1$ and $a=-1/2$. 

Performing independent variations of the action (\ref{eq:action}) with respect to metric and connection leads to
\begin{eqnarray}
f_R R_{\mu\nu}-\frac{f}{2}g_{\mu\nu}+2f_QR_{\mu\alpha}{R^\alpha}_\nu &=& \kappa^2 T_{\mu\nu}\label{eq:met-varX}\\ \nonumber \\
\nabla_{\beta}\left[\sqrt{-g}\left(f_R g^{\mu\nu}+2f_Q R^{\mu\nu}\right)\right]&=&0 
 \label{eq:con-varX}\\\nonumber \\
  f_R \equiv \dfrac{df}{dR} \quad , \quad f_Q \equiv \dfrac{df}{dQ}\nonumber
\end{eqnarray}
In order to solve (\ref{eq:met-varX}) and (\ref{eq:con-varX}) we introduce the matrix  $\hat{P}$ (whose components are  ${P_\mu}^\nu\equiv R_{\mu\alpha}g^{\alpha\nu}$), which allows us to express them as ( where $\hat{I}$ is the identity matrix)
\begin{equation}
2f_Q\hat{P}^2+f_R \hat{P}-\frac{f}{2}\hat{I} = \kappa^2 \hat{T}
\label{eq:met-varRQ22}
\end{equation}
\begin{equation}
\nabla_{\beta}[\sqrt{-g} g^{\mu\alpha} {\Sigma_\alpha}^\nu]=0 
 \label{eq:met-varRQ2}
\end{equation}
where $\hat{T}$, the matrix representation of ${T_\mu}^\nu$, and ${\Sigma_\alpha}^{\nu}$ are given by
\begin{equation}\label{eq:Tmn-EM}
{T_\mu}^\nu=\frac{1}{8\pi} \left( \begin{array}{cc}
 (\varphi-2X\varphi_X) \hat{I} & \hat{0}  \\
\hat{0} & \varphi \hat{I} \\
\end{array} \right) \quad , \quad {\Sigma_\alpha}^{\nu}=\left(f_R \delta_{\alpha}^{\nu} +2f_Q {P_\alpha}^{\nu}\right)
\end{equation}
We can now transform to what is called Einstein frame by introducing a new symmetric rank-two tensor $h^{\mu\nu}$ satisfying
\begin{equation} \label{eq:auxmetric}
\nabla_{\beta}[\sqrt{-g} g^{\mu\alpha} {\Sigma_\alpha}^\nu]=\nabla_{\beta}[\sqrt{-h} h^{\mu\nu}]=0 \ .
\end{equation}
which leads to the following solution
\begin{equation} \label{eq:h-g}
h^{\mu\nu}=\frac{g^{\mu\alpha}{\Sigma_{\alpha}}^\nu}{\sqrt{\det \hat{\Sigma}}} \ , \quad
h_{\mu\nu}=\left(\sqrt{\det \hat{\Sigma}}\right){\Sigma_{\mu}}^{\alpha}g_{\alpha\nu} \ .
\end{equation}
Using the definition of ${\Sigma_\mu}^\nu$ and the relations (\ref{eq:h-g}), allows us to express the metric field equations using $h_{\mu\nu}$ in a compact form
\begin{equation} \label{eq:fieldequations}
{R_{\mu}}^{\nu}(h)=\frac{1}{\sqrt{\det \hat{\Sigma}}}\left(\frac{f}{2}{\delta_{\mu}}^{\nu}+ \kappa^2 {T_{\mu}}^{\nu} \right) \ .
\end{equation}
Next, to find the explicit form of $\hat{P}$ for our problem, which is necessary to obtain $\hat{\Sigma}$, we write equation (\ref{eq:met-varRQ22}) as
\begin{equation}
2f_Q\left(\hat{P}+\frac{f_R}{4f_Q}\hat{I}\right)^2= \left(
\begin{array}{cc}
\lambda_-^2\hat{I} & \hat{0}  \\
\hat{0} & \lambda_+^2\hat{I} \\
\end{array} \right),  \label{eq:P}
\end{equation}
where
\begin{equation}
 \lambda_{+}^2=\frac{1}{2}\left(f+\frac{f_R^2}{4f_Q}+2k^2 T_{\theta}^{\theta}\right) \label{eq:lambda+} \quad , \quad \lambda_{-}^2=\frac{1}{2} \left(f+\frac{f_R^2}{4f_Q} +2k^2 T_{t}^{t}\right) 
\nonumber
\end{equation}
Taking the square root in (\ref{eq:P}) and demanding agreement with GR in the low curvature regime (where $f_R\to 1$ and $f_Q\to 0$) it follows that the matrix $\hat{\Sigma}$ is given by
\begin{equation} \label{eq:sigma-matrix}
\hat{\Sigma}=\frac{f_R}{2}\hat{I}+\sqrt{2f_Q} \left(
\begin{array}{cc}
\lambda_- \hat{I}& \hat{0} \\
\hat{0} & \lambda_+ \hat{I} \\
\end{array} \right)= \left(
\begin{array}{cc}
\sigma_- \hat{I}& \hat{0} \\
\hat{0} & \sigma_+\hat{I} \\
\end{array}
\right),
\end{equation}
where 
\begin{equation}
\sigma_{+}=1-2\kappa^2l_P^2 T_{\theta}^{\theta} \quad , \quad \sigma_{-}= 1-2\kappa^2l_P^2 T_{t}^{t}\label{eq:sigma}
\nonumber
\end{equation}
Gathering all these elements we obtain the field equations
for our model
\begin{equation}\label{eq:Rmn}
{R_{\mu}}^{\nu}(h)=\frac{1}{2 \sigma_{+} \sigma_{-}} \left(
\begin{array}{cc}
(f+2\kappa^2 T_t^t)\hat{I} & \hat{0} \\
\hat{0} & (f+2\kappa^2 T_{\theta}^{\theta}) \hat{I} \\
\end{array}
\right).
\end{equation}


To solve the field equations (\ref{eq:Rmn}) we introduce two different line elements in Schwarzschild-like coordinates, one associated to the physical metric $g_{\mu\nu}$ and another associated to the auxiliary metric $h_{\mu\nu}$, respectively;
\begin{equation}\label{eq:ds2g}
ds^2= g_{tt}dt^2+g_{rr}dr^2+r^2 d\Omega^2 \qquad d\tilde{s}^2= h_{tt}dt^2+h_{rr}dr^2+\tilde{r}^2 d\Omega^2 
\end{equation}
The relation between these two line elements is obtained via $g_{\mu\nu}={\Sigma_\mu}^{\alpha}h_{\alpha\nu}/\sqrt{\det \Sigma}$, which implies that $\tilde{r}^2 =r^2 \sigma_{-}$. Using the ansatzes: 
\begin{equation}
h_{tt}=-A(\tilde{r})e^{2\psi(\tilde{r})},\quad h_{\tilde{r}\tilde{r}}=1/A(\tilde{r}),\quad A(\tilde{r})=1-\dfrac{2M(\tilde{r})}{\tilde{r}}
\end{equation}
one finds that $\psi=0$, and that $M(r)$ satisfies
\begin{equation} \label{eq:mass-r}
 \dfrac{M(r)}{M_0}=1+\delta_1 G(r)
\end{equation}
where $M_0$ and $\delta_1$ are constants, and the function $G(r)$  satisfies
\begin{equation} \label{eq:Gzz}
\frac{dG}{dr}=\frac{r^2 \sigma_{-}^{1/2}}{\sigma_+}\left(1+\frac{r\sigma_{-,r}}{2\sigma_-}\right)\left(\frac{4\pi}{\kappa^2}f+\varphi\right) \ .
\end{equation}
With all these elements we arrive to the final expression for the physical metric components in (\ref{eq:ds2g}) as
\begin{eqnarray} 
g_{tt}=-\frac{A(r)}{\sigma_{+}(r)} &;& g_{rr}=\frac{\sigma_{-}(r) }{\sigma_{+}(r)A(r)} \left(1+\frac{r\sigma_{-,r}}{2\sigma_{-}(r)} \right)^2\label{eq:A1}\\
A(r)&=&1-\frac{1+\delta_1 G(r)}{\delta_2 r \sigma_{-}(r)^{1/2}} \label{eq:A2} \ ,
\end{eqnarray}
where $\delta_1$ and $\delta_2$ are constants. 


\subsection{Solutions With Wormhole Topology}
In the case of (\ref{eq:lagrangian-confining}) we find ( to simplify the notation and the physical meaning of the solution we define a new dimensionless variable $z=r/r_c$, with $r_c\equiv \sqrt{l_P r_q}$, and a new parameter $\delta_3=l_P/r_g$) 
 that the function $G_z$ is explicitly obtained from (\ref{eq:Gzz}) as
\begin{equation} \label{eq:Gz}
\frac{dG}{dz}=\left(1+\delta_3 z^2 \right)^2\left(\frac{z^4(1-\delta_3^2)+1}{z^4 \sqrt{z^4(1-\delta_3^2) - 2 \delta_3 z^2 -1}}\right)
\end{equation}
To integrate $G_z$ we note that this function is only defined for $z>z_c=\frac{1}{\sqrt{1-\delta_3}}$, for which the denominator of (\ref{eq:Gz}) vanishes. This corresponds to the point where the function $\sigma_{-}$ vanishes and is intimately related to the existence of a wormhole. In a sense, the $z=z_c$ surface defines a core that replaces the usual point-like singularity of GR. Taking into account the existence of this core we can exactly integrate the function $G_z$. 

To do this we find it useful to introduce the following change of variable
\begin{equation}\label{eq:z2x}
z=\frac{x}{\sqrt{1-\delta_3}}=x\sqrt{\frac{1+\chi}{{2}}} \ ,
\end{equation}
where we have introduced the constant $\chi$ through $\delta_3=\frac{\chi-1}{\chi+1}$, with $\chi\ge 1$ and such that $\chi \rightarrow 1$ implies $\delta_3 \rightarrow 0$ (the Maxwell limit). With these definitions the function $G_x=G_z dz/dx$ becomes
\begin{equation}\label{eq:Gx3}
G_x= \frac{\left((\chi -1) x^2+2\right)^2 \left(\chi  x^4+1\right)}{\sqrt{2} (\chi +1)^{3/2} x^4 \sqrt{\left(x^2-1\right) \left(\chi  x^2+1\right)}}
\end{equation}
The integration of $G_x$ is know analytical and when it is performed the solutions is given in terms of  elliptic integrals of the second and first kind.
\begin{enumerate}
\item \textbf{Geometry and topology as $x\to 1$}

The behaviour of the metric near the core, $x_c=1$, is obtained from (\ref{eq:Gx3}) after we perform a series expansion of $g_{tt}$ around $x=1$ ( ${\delta_c^\chi}$ is a constant function of $\chi$)
\begin{equation}\label{eq:gtt_series}
g_{tt} \simeq  \frac{(\chi +1) ({\delta_c^\chi}-\delta_1)}{8 \delta_2 {\delta_c^\chi} \sqrt{x-1}} + \frac{(\chi +1) (\delta_1 (\chi +1)-2 \delta_2)}{8 \delta_2}+\ldots 
\end{equation}
While this function is in general divergent as $x \rightarrow 1$, for $\delta_1={\delta_c^\chi}$ we find a finite result
\begin{equation}
g_{tt} \simeq \frac{(\chi +1) ({\delta_c^\chi} ( \chi +1)-2 \delta_2)}{8 \delta_2} + O(x-1) \ . \label{eq:gtt}
\end{equation}
This result indicates that the line element near $x=1$ describes a Minkowskian-like region. 
As a result, curvature invariants such as $R(g)$, $Q(g)$ and the Kretschmann scalar,  
turn out to be finite at $x=1$ for arbitrary $\chi$ when $\delta_1={\delta_c^\chi}$.
\item \textbf{de Sitter behaviour when $x\gg 1$}

Expanding $g_{tt}$ in $x \gg 1$, restoring the usual notation, and keeping only second-order terms, we find
\begin{eqnarray} \label{eq:asymp}
g_{tt} &\simeq &-\left(1-\frac{r_q}{r_g}+\frac{l_P^2}{r_g^2}\right)+\frac{r_S }{r}-\frac{r_q^2}{2 r^2} +\frac{r^2}{6 r_g^2}  +O\left(\frac{1}{r^3}\right) \ .
\end{eqnarray}
This expression is not exactly a Reissner-Nordstr\"{o}m-de-sitter solution because of the constant added terms $-\frac{r_q}{r_g}+\frac{l_P^2}{r_g^2}$, where we have kept the minimum number of terms containing both ultraviolet corrections due to the gravitational Lagrangian, which manifest the existence of a wormhole structure, and  infrared corrections coming from the matter sector, which are responsible for the asymptotically de Sitter behavior. Let us recall that no bare cosmological constant  $\Lambda_B$ was included in the starting action (\ref{eq:action}).
\end{enumerate}
Thus we find the existence of a set of solutions for which the curvature invariants are smooth everywhere and the space-time possesses a wormhole structure, representing an object that can be naturally interpreted in terms of Wheeler's geon \cite{Wheeler, lor13}{.} These solutions admit an exact analytical expression and arise naturally from the nonlinear electromagnetic field rather than requiring exotic matter to generate  a pre-designed wormhole geometry, such the ones we have found when we coupled the non-linear gauge filed to gravity via light/time-like branes.

\newpage
\section{Non-Linear Maxwell Term, Dilaton and $R^2$-Gravity}
We now consider the coupling of $f(R)= R + \alpha R^2$ gravity (possibly
with a bare cosmological constant $\Lambda_0$) to a ``dilaton'' field , $\phi$, and
the nonlinear gauge field system containing $\sqrt{-F^2}$ where in this case the relevant is given by ( again, we work within the Palatini formalism):
\begin{equation}
S = \int d^4 x \sqrt{-g} \left[ \frac{1}{16\pi} 
\left( f\left( R(g,\Gamma)\right) - 2\Lambda_0 \right) + L(F^2(g)) + L_D (\phi,g) \right]
\label{eq:f-gravity+GG+D} 
\end{equation}
where
\begin{equation}
f\left( R(g,\Gamma)\right) = R(g,\Gamma) + \alpha R^2(g,\Gamma) \quad ,\quad 
R(g,\Gamma) = R_{\mu\nu}(\Gamma) g^{\mu\nu}
\label{eq:f-gravity}
\end{equation}
\begin{equation}
L(F^2(g)) = - \frac{1}{4e^2} F^2(g) - \frac{f_0}{2} \sqrt{- F^2(g)}
\label{eq:GG-g}
\end{equation}
\begin{equation}
F^2(g) \equiv F_{\alpha\beta} F_{\mu\nu} g^{\alpha\mu} g^{\beta\nu} \quad ,\quad
F_{\mu\nu} = \partial_\mu A_\nu - \partial_\nu A_\mu
\label{eq:F2-g}
\end{equation}
\begin{equation}
L_D (\phi,g) = -\dfrac{1}{2} g^{\mu\nu}\partial_\mu \phi \partial_\nu \phi - V(\phi) 
\label{eq:L-dilaton}
\end{equation}
$R_{\mu\nu}(\Gamma)$ is the Ricci curvature in the first order (Palatini) formalism, 
\textsl{i.e.}, the spacetime metric $g_{\mu\nu}$ and the affine connection 
$\Gamma^\mu_{\nu\alpha}$ are \textsl{a priori} independent variables. The equations of motion resulting from the action (\ref{eq:f-gravity+GG+D}) read:
\begin{equation}
R_{\mu\nu}(\Gamma) = \frac{1}{f^{'}_R}\left[8\pi T_{\mu\nu} + 
\dfrac{1}{2} f\left( R(g,\Gamma)\right) g_{\mu\nu}\right] 
\quad , \quad f^{'}_R \equiv \frac{df(R)}{dR} = 1 + 2\alpha R(g,\Gamma)
\label{eq:g-eqs}
\end{equation}
\begin{equation}
\nabla_\lambda \left(\sqrt{-g} f^{'}_R g^{\mu\nu}\right)  = 0
\label{eq:gamma-eqs}
\end{equation}
\begin{equation}
\partial_\nu \left(\sqrt{-g} \left[ 1/e^2 - \frac{f_0}{\sqrt{-F^2(g)}}
\right] F_{\alpha\beta} g^{\mu\alpha} g^{\nu\beta}\right)=0
\label{eq:GG-eqs-R2}
\end{equation}
And the total energy-momentum tensor is given by:
\begin{eqnarray}
T_{\mu\nu} = 
\left[ L(F^2(g))+L_D (\phi,g)-\frac{1}{8\pi}\Lambda_0\right] g_{\mu\nu} 
\nonumber \\
+ \left(1/e^2 - \frac{f_0}{\sqrt{-F^2(g)}}\right)
F_{\mu\alpha} F_{\nu\beta} g^{\alpha\beta} + \partial_\mu \phi \partial_\nu \phi
\label{eq:T-total}
\end{eqnarray}
\subsection{Einstein-Frame Formalism and Action}
Equation (\ref{eq:gamma-eqs}) leads to the relation $\nabla_\lambda \left( f^{'}_R g_{\mu\nu}\right)=0$
and thus it implies transition to the ``physical'' Einstein-frame metrics 
$h_{\mu\nu}$ via conformal rescaling of the original metric $g_{\mu\nu}$, this is similar to what we have done in the previous section, only here we have $f^{'}_R g_{\mu\nu} =  h_{\mu\nu}$. Using this relation, the $R^2$-gravity equations of motion (\ref{eq:g-eqs}) can be
rewritten in the form of {\em standard} Einstein equations:
\begin{equation}
R^\mu_\nu (h) = 8\pi \left({T_{eff}}^\mu_\nu (h) - \dfrac{1}{2} \delta^\mu_\nu {T_{eff}}^\lambda_\lambda (h)\right)
\label{eq:einstein-h-eqs}
\end{equation}
with effective energy-momentum tensor of the following form:
\begin{equation}
{T_{eff}}_{\mu\nu} (h) = h_{\mu\nu} L_{eff} (h) 
- 2\dfrac{\partial L_{eff}}{\partial h^{\mu\nu}}
\label{eq:T-h-eff}
\end{equation}
and the effective Einstein-frame matter Lagrangian is given by 
\begin{equation}
L_{eff} (h) = - \frac{1}{4 e_{eff}^2 (\phi)} F^2(h) 
- \dfrac{1}{2} f_{eff} (\phi) \sqrt{- F^2(h)} -V_{eff}(\phi)
\label{eq:L-eff-h}
\end{equation}
where 
\begin{equation}
V_{eff}(\phi)\equiv -\frac{X(\phi,h)\left(1+16\pi\alpha X(\phi,h)\right) - V(\phi) -\Lambda_0 / 8\pi
}{1+8\alpha\left( 8\pi V(\phi)+\Lambda_0\right)}
\end{equation}
$X(\phi,h) \equiv -\dfrac{1}{2} h^{\mu\nu}\partial_\mu \phi \partial_\nu \phi$ is the kinetic energy of the dilaton field $\phi$, and the dynamical $\phi$-dependent couplings are given by:
\begin{eqnarray}
\frac{1}{e_{eff}^2 (\phi)} = \frac{1}{e^2} + 
\frac{16\pi\alpha f_0^2}{1 + 8\alpha \left(8\pi V(\phi) + \Lambda_0 \right)}
\label{eq:e-eff} \\
f_{eff}(\phi)=f_0 \frac{1+32\pi\alpha X(\phi,h)}{1 + 8\alpha\left(8\pi V(\phi)+\Lambda_0\right)}
\label{eq:f-eff}
\end{eqnarray}
Thus, all equations of motion of the original $R^2$-gravity system 
(\ref{eq:f-gravity+GG+D})--(\ref{eq:L-dilaton}) can be equivalently derived from the 
following Einstein/nonlinear-gauge-field/dilaton action:
\begin{equation}
S_{eff} = \int d^4 x \sqrt{-h} \left[ \frac{R(h)}{16\pi} 
+ L_{eff} (h)\right]
\label{eq:einstein-frame-action}
\end{equation}
One should notice that even if the ordinary kinetic Maxwell term $-\frac{1}{4}F^2$ is absent in the original system ($e^2 \to \infty$ in (\ref{eq:GG-g})), such term is nevertheless {\em dynamically generated} in the Einstein-frame action (\ref{eq:L-eff-h})--(\ref{eq:einstein-frame-action}), this feature is a {\em combined effect} of the $\alpha R^2$ term and the non-linear Maxwell term $-\dfrac{f_0}{2}\sqrt{-F^2}$.
\subsection{Confining-Deconfinement phases}
In what follows we consider constant ``dilaton'' $\phi$ (thus the dilaton kinetic term is equal to zero, $X(\phi,h)=0$) extremizing the effective Lagrangian (\ref{eq:L-eff-h}), in fact, one should notice that the dynamical couplings and effective potential are extremized 
{\em simultaneously}, this is an explicit realization of ``least coupling principle'' of Damour-Polyakov \cite{damour-polyakov}{:}
\begin{equation}
\dfrac{\partial f_{eff}}{\partial\phi} = - 64\pi\alpha f_0 \dfrac{\partial V_{eff}}{\partial \phi}
\quad ,\quad \dfrac{}{\partial\phi}\frac{1}{e_{eff}^2} =
-(32\pi\alpha f_0)^2 \dfrac{\partial V_{\rm eff}}{\partial\phi} 
\quad \to \dfrac{\partial L_{eff}}{\partial\phi} \sim \dfrac{\partial V_{eff}}{\partial\phi}
\label{eq:f-e-extremize}
\end{equation}
Therefore at the extremum of $L_{eff}(h)$, $\phi$ must satisfy the following relation:
\begin{equation}
\dfrac{\partial V_{eff}}{\partial\phi} = 
\frac{V^{'}(\phi)}{\left[ 1+8\alpha\left(\kappa^2 V(\phi)+\Lambda_0\right)\right]^2} = 0
\label{eq:V-extremum}
\end{equation}
There are two generic distinct cases:
\begin{itemize}
\item[(a)] {\em Confining phase}: \\ Equation (\ref{eq:V-extremum}) is satisfied for some 
finite-value $\phi_0$ extremizing the original potential $V(\phi)$: 
\begin{equation}
V^{'}(\phi_0) = 0
\end{equation}
\item[(b)] {\em Deconfinement phase}:\\ For polynomial or exponentially 
growing original $V(\phi)$, so that $V(\phi) \to \infty$ when $\phi \to \infty$, 
we have:
\begin{equation}
\dfrac{\partial V_{eff}}{\partial\phi} \to 0 \quad ,\quad 
V_{eff} (\phi) \to \frac{1}{64\pi\alpha} = {const} \quad \text{when} \quad
\phi \to \infty
\label{eq:flat-region}
\end{equation}
meaning for sufficiently large values of $\phi$ we find a ``flat region''
in $V_{eff}$.
\end{itemize}
This ``flat region'' triggers a {\em transition from 
confinement to deconfinement dynamics}. Namely, in the ``flat-region'' case ($V(\phi) \to \infty$) we have:
\begin{equation}
f_{eff} \to 0 \quad ,\quad e^2_{eff} \to e^2
\label{eq:deconfine}
\end{equation}
and the effective gauge field Lagrangian, $L_{eff}$, reduces to the ordinary
\textsl{non-confining} one (the ``square-root'' term $\sqrt{-F^2}$ vanishes):
\begin{equation}
L^{(0)}_{eff} = -\frac{1}{4e^2} F^2(h) - \frac{1}{64\pi\alpha}
\label{eq:L-eff-h-0}
\end{equation}
with an {\em induced} cosmological constant $\Lambda_{eff} = 1/8\alpha$, which is
{\em completely independent} of the bare cosmological constant $\Lambda_0$.
\subsection{Generalized Solutions}
Within the physical ``Einstein''-frame in the confining phase 
($V^{'}(\phi_0) = 0 $, $\phi_0 =\mathrm{finite}$) we find generalizations of the previously obtained solutions:
\begin{enumerate}
\item Reissner-Nordstr{\"o}m-({\em anti}-)de-Sitter type black holes, in
particular,  non-standard Reissner-Nordstr{\"o}m type with non-flat ``hedgehog''
asymptotics, generalizing solutions (\ref{eq:cornell-sol}) and (\ref{eq:RN-dS+const-electr}) in the ordinary Einstein-gravity case, where now the effective cosmological constant and the vacuum constant radial electric field read:
\begin{equation}
\Lambda_{eff}(\phi_0) = \frac{\Lambda_0 +8\pi V(\phi_0)+2\pi e^2 f^2_0}{
1+8\alpha\left(\Lambda_0 +8\pi V(\phi_0)+2\pi e^2 f^2_0\right)} 
\label{eq:h-CC-eff}
\end{equation}
\begin{equation}
|\vec{E}_{vac}| = \left(\frac{1}{e^2} + 
\frac{16\pi\alpha f_0^2}{1 + 8\alpha\left(8\pi V(\phi_0) + \Lambda_0 \right)}\right)^{-1}
\frac{f_0/\sqrt{2}}{1+8\alpha\left(8\pi V(\phi_0)+\Lambda_0\right)} 
\label{eq:vacuum-radial}
\end{equation}
\item Levi-Civita-Bertotti-Robinson type ``tubelike'' spacetimes generalizing (\ref{eq:gen-BR-metric})--(\ref{eq:dS2}),
where now (using short-hand notation $\Lambda(\phi_0)\equiv 8\pi V(\phi_0) + \Lambda_0$):
\begin{equation}
\frac{1}{r_0^2} = \frac{4\pi}{1+8\alpha\Lambda(\phi_0)}\left[
\left(1+8\alpha\left(\Lambda(\phi_0)+2\pi f_0^2\right)\right) \vec{E}^2 +
\frac{1}{4\pi}\Lambda(\phi_0)\right]
\label{eq:r0-eq}
\end{equation}
\end{enumerate}

\newpage
\section{TMT Approach to Bags and Confinement}
\subsection{TMT Fundamentals}

``Two Measure Theory" is a generally coordinate invariant theory where all the difference from the standard field theory in curved space-time consists of the following assumptions:

\begin{enumerate}
\item The effective action includes two Lagrangians, $ L_{1}$ and $L_{2}$, and two
measures of integration $\sqrt{-g}$ and $\Phi$. One is the usual measure of integration $\sqrt{-g}$
in the 4-dimensional space-time manifold equipped with the metric
 $g_{\mu\nu}$. Another  is the new measure of integration $\Phi$ in the same
4-dimensional space-time manifold; 
\begin{equation}\label{eq:eg1e6}
S = \int L_{1} \sqrt{-g}d^{4}x + \int L_{2} \Phi  d^{4} x    
\end{equation}
The measure  $\Phi$ being  a
scalar density and a total derivative may be defined by means of  four scalar fields $\varphi_{a}$ ($a=1,2,3,4$)\footnote{
$\Phi$ can also be defined by means of a totally antisymmetric three index field
$A_{\alpha\beta\gamma}$
\begin{equation}
\Phi
=\varepsilon^{\mu\nu\alpha\beta}\partial_{\mu}A_{\nu\alpha\beta}.
\nonumber
\end{equation}
}
\begin{equation}
\Phi
=\varepsilon^{\mu\nu\alpha\beta}\varepsilon_{abcd}\partial_{\mu}\varphi_{a}
\partial_{\nu}\varphi_{b}\partial_{\alpha}\varphi_{c}
\partial_{\beta}\varphi_{d}.
\label{eq:eg12}
\end{equation}

\item It is assumed that the Lagrangians $ L_{1}$ and $L_{2}$ are
functions of all matter fields, the metric, the connection  (or
spin-connection )
 but not of the
"measure fields" ($\varphi_{a}$ or $A_{\alpha\beta\gamma}$).

\item The metric, connection (or vierbein and spin-connection) and the
{\it measure fields} ($\varphi_{a}$ or $A_{\alpha\beta\gamma}$)
are independent dynamical variables. All the relations between
them are the results of the equations of motion (the
independence of the metric and the connection means that we
work in the Palatini formalism)
\end{enumerate}
Globally scale invariant realization of such kind of theories, which require the introduction of a dilaton field $\phi$, with a generalization to include an "$R^{2}$ term" is given by the action 
\begin{equation}\label{eq:eg1e10}
L_{1} = U(\phi) + \epsilon R(\Gamma, g)^{2} 
\end{equation}
\begin{equation}\label{eq:eg1e11}
L_{2} = \frac{-1}{\kappa} R(\Gamma, g) + \frac{1}{2} g^{\mu\nu}
\partial_{\mu} \phi \partial_{\nu} \phi - V(\phi)
\end{equation}
\begin{equation}\label{eq:eg1e12}
R(\Gamma,g) =  g^{\mu\nu}  R_{\mu\nu} (\Gamma) , R_{\mu\nu}
(\Gamma) = R^{\lambda}_{\mu\nu\lambda}
\end{equation}
\begin{equation}\label{eq:eg1e13}
R^{\lambda}_{\mu\nu\sigma} (\Gamma) = \Gamma^{\lambda}_
{\mu\nu,\sigma} - \Gamma^{\lambda}_{\mu\sigma,\nu} +
\Gamma^{\lambda}_{\alpha\sigma}  \Gamma^{\alpha}_{\mu\nu} -
\Gamma^{\lambda}_{\alpha\nu} \Gamma^{\alpha}_{\mu\sigma}.
\end{equation}
global scale invariance is satisfied if 
\begin{equation}\label{eq:eg1e19} 
V(\phi) = f_{1}  e^{\alpha\phi},  U(\phi) =  f_{2}
e^{2\alpha\phi}
\end{equation}
where $ f_{1}, f_{2},\alpha $ are constants, while for $U(\phi)=V(\phi)=0$ we have local conformal invariance. 

A particularly interesting equation is the one that arises from the $\varphi_{a}$
fields variation, this yields $L_2 = M$, where $M$ is a constant that spontaneously breaks scale invariance.

The Einstein frame for this case, as we already observed, is a redefinition of the metric by a conformal factor, and is defined as 
\begin{equation}\label{eq:eg1e47}
\overline{g}_{\mu\nu} = (\chi -2\kappa \epsilon R) g_{\mu\nu}
\end{equation}
where $\chi$ is the ratio between the two measures, $\chi =\dfrac{\Phi}{\sqrt{-g}}$, and is determined from the consistency of the equations to be $\chi = \dfrac{2U(\phi)}{M+V(\phi)}$. One can then find an effective action, where we can use this Einstein frame metric, which is given by 
\begin{equation}
S_{eff}=\int\sqrt{-\overline{g}}d^{4}x\left[-\frac{1}{\kappa}\overline{R}(\overline{g})
+p\left(\phi,R\right)\right] \label{eq:eg1k-eff}
\end{equation}
\begin{equation}
 p = \frac{\chi}{\chi -2 \kappa \epsilon R}X - V_{eff} \quad , \quad X = \frac{1}{2} \overline{g}^{\mu\nu}\partial_{\mu} \phi \partial_{\nu} \phi \quad , \quad  V_{eff}  = \frac{\epsilon R^{2} + U}{(\chi -2 \kappa \epsilon R)^{2} }
\end{equation}
$\overline{R}$ is the Riemannian curvature scalar built out of the bar metric, $R$ on the other hand is the non Riemannian curvature scalar defined in terms of the connection and the original metric, which turns out to be  
\begin{equation}
R = \dfrac{-\kappa (V+M) +\frac{\kappa}{2} \overline{g}^{\mu\nu}\partial_{\mu} \phi \partial_{\nu} \phi \chi}
{1 + \kappa ^{2} \epsilon \overline{g}^{\mu\nu}\partial_{\mu} \phi \partial_{\nu} \phi}
\end{equation}
Introducing this $R$ into the expression for $V_{eff}$ and considering a constant ``dilaton" field $\phi$ we find that $ V_{eff}$ has two flat
regions as a consequence of the s.s.b. of the scale symmetry (that is of considering  $M \neq 0$ ).  
\subsection{Softly Broken Conformally Invariant TMT Model}
Now we turn to incorporating the gauge field terms $ \sqrt { - F_{\mu \nu }^{a}\, F^{a\mu \nu } }$ and $- F_{\mu \nu }^{a}\, F^{a\mu \nu } $. We first notice that the $ \sqrt { - F_{\mu \nu }^{a}\, F^{a\mu \nu } }$ term respects conformal symmetry if coupled to the new measure $\Phi$. To see this we note that under conformal  transformation 
\begin{equation}\label{eq:eg1e14}
g_{\mu\nu}  \rightarrow   \Omega(x)  g_{\mu\nu} \quad , \quad \Phi \rightarrow \Phi^{\prime} = J(x) \Phi   
\end{equation}
where $J(x)$  is the Jacobian of the transformation of the $\varphi_{a}$ fields.
This will be a symmetry in the case $U=V=0$ if 
\begin{equation}\label{eq:eg1e17}
\Omega = J
\end{equation}
since 
$\sqrt { - F_{\mu \nu }^{a}\, F^{a\mu \nu }}$ $=\sqrt { - F_{\mu \nu }^{a}\, F^{a}_{\alpha\beta}g^{\mu\alpha}g^{\nu\beta}}$, then according to (\ref{eq:eg1e14}) $\sqrt { - F_{\mu \nu }^{a}\, F^{a\mu \nu }}\rightarrow \Omega^{-1}\sqrt { - F_{\mu \nu }^{a}\, F^{a\mu \nu }}$ if $\Omega>0$
and $\Phi \rightarrow  J \Phi= \Omega \Phi$, so that $\Phi\sqrt { - F_{\mu \nu }^{a}\, F^{a\mu \nu }}$ is invariant. A similar situation happens with a mass term for the gluon, $A_{\mu }^{a}\, A^{a}_{\alpha}g^{\mu\alpha}$ in TMT, this can be a conformally invariant
if it goes multiplied with the measure $\Phi$. Furthermore, conformal invariance implies that a term proportional to
 $F_{\mu \nu }^{a}F^{a}_{\alpha\beta}g^{\mu\alpha}g^{\nu\beta}$ has to appear multiplied by the measure $\sqrt{-g}$, since $\sqrt{-g}F_{\mu \nu }^{a}F^{a}_{\alpha\beta}g^{\mu\alpha}g^{\nu\beta}$ is invariant under conformal transformations of the metric. Therefore a softly broken conformal invariant model will have the form 
\begin{equation}\label{eq:eg1finalaction}
S = S_L + S_{R^2} - \frac  {1}{4}\int d^4x\sqrt{-g}F_{\mu \nu }^{a}F^{a}_{\alpha\beta}g^{\mu\alpha}g^{\nu\beta} 
+\frac{N}{2}\int d^4x \Phi\sqrt{-F_{\mu \nu }^{a}F^{a}_{\alpha\beta}g^{\mu\alpha}g^{\nu\beta}}
\end{equation}
This action does not contain mass terms for the gluons, the consequences of having both a mass term and a confinement term have been explored in [\refcite{eg1interplay}] where it was shown that in such a case confinement is lost in favor of a Coulomb like behavior.
\subsection{Bag Dynamics in Softly Broken Conformally Invariant TMT Model}
Let us proceed now to describe the consequences of the action (\ref{eq:eg1finalaction}). 
The steps to follow are the same as in the case where we did not have gauge fields (vary the action with respect to $\Gamma$, $g_{\mu\nu}$ and $\Phi$, find the constraints and make the transition to the Einstein frame).

Now, for the case where gauge fields are included in the way described by (\ref{eq:eg1finalaction}), 
all the equations of motion in the Einstein frame will be correctly described by 
\begin{equation}\label{eq:eg1finalbagmodel}
S_{eff}=\int\sqrt{-\bar{g}}d^{4}x\left[-\frac{1}{\kappa}\bar{R}(\bar{g})
+p\left(\phi,R,X,F_{\mu \nu }^{a}, \bar{g}^{\alpha\beta}\right)\right] 
\end{equation}
\begin{equation}\label{eq:eg1newp}
 p = \frac{\chi}{\chi -2 \kappa \epsilon R} \left[X
+\frac{N}{2}\sqrt{-F_{\mu \nu }^{a}F^{a}_{\alpha\beta}\bar{g}^{\mu\alpha}\bar{g}^{\nu\beta}}\right] - \frac{1}{4}F_{\mu \nu }^{a}F^{a}_{\alpha\beta}\bar{g}^{\mu\alpha}\bar{g}^{\nu\beta} 
 - V_{eff}
\end{equation}
\begin{equation}\label{eq:eg1newV}
 V_{eff}  = \frac{\epsilon R^{2} + U}{(\chi -2 \kappa \epsilon R)^{2} }
\end{equation}
where 
\begin{equation}\label{eq:eg1chi3}
\chi = \frac{2U(\phi)}{M+V(\phi)}
\end{equation}
One interesting fact is that the terms that enter the constraint that determines $\chi$ are only those that break the conformal invariance and  the constant of integration $M$.  Since the new terms involving the gauge fields do not break the conformal invariance the relevant terms that violate this symmetry are only the $U$ and $V$ terms and the constraint remains the same.

Again we notice that $\bar{R}$ and $R$ are different objects, the $\bar{R}$ is the Riemannian curvature scalar in the Einstein frame,
while $R$ is a different object which now is given by
\begin{equation}\label{eq:eg1newR}
R = \frac{-\kappa (V+M) +\kappa \chi \left( X +\frac{N}{2}\sqrt{-F_{\mu \nu }^{a}F^{a}_{\alpha\beta}\bar{g}^{\mu\alpha}\bar{g}^{\nu\beta}}\right)}
{1 + 2\kappa ^{2} \epsilon \left(X +\frac{N}{2}\sqrt{-F_{\mu \nu }^{a}F^{a}_{\alpha\beta}\bar{g}^{\mu\alpha}\bar{g}^{\nu\beta}}\right) }
\end{equation}
\subsection{Regular gauge field dynamics inside the bags}
From (\ref{eq:eg1finalbagmodel}), (\ref{eq:eg1newp}) and (\ref{eq:eg1chi3}), we see that the $N$ term, responsible for the confining gauge dynamics, gets dressed in the Einstein frame effective action by the factor $\dfrac{\chi}{\chi -2 \kappa \epsilon R}$. As we consider regions inside the bags, where $\phi \rightarrow -\infty$, we see  that $\chi$ approaches zero for 
$M \neq 0$, therefore for the case $\epsilon \neq 0$
the $N$ term inside the bags disappears. Therefore the only term providing gauge field dynamics is the standard term $-\frac{1}{4}F_{\mu \nu }^{a}F^{a}_{\alpha\beta}\bar{g}^{\mu\alpha}\bar{g}^{\nu\beta} $. 

In this same limit and with the same conditions, using only that as $\phi \rightarrow -\infty$, $U \rightarrow 0$ and $\chi \rightarrow 0 $, we see that still, in the more complicated theory with gauge fields the same bag constant  $V_{eff} \rightarrow  \frac{1}{4\epsilon \kappa^{2}}$ is obtained, so $V_{eff}$ does not contribute to the gauge field equations of motion, but does provide the Bag constant.


\subsection{Confining gauge field effective action outside the bags} 
Take now the opposite limit, $\phi \rightarrow +\infty$, the scalar field outside the bag is pushed to large values of $\phi$, since the absolute minimum of the effective potential is found for such values, then   confining dynamics appears,
\begin{equation}\label{eq:eg1Vlimit}
 V_{eff}   \rightarrow C+4B \left[\,X
+\frac{N}{2}\sqrt{-F_{\mu \nu }^{a}F^{a}_{\alpha\beta}\bar{g}^{\mu\alpha}\bar{g}^{\nu\beta}}\ \right]^2
\end{equation}
and
 $ \dfrac{\chi}{\chi -2 \kappa \epsilon R}\left[\,X
+\frac{N}{2}\sqrt{-F_{\mu \nu }^{a}F^{a}_{\alpha\beta}\bar{g}^{\mu\alpha}\bar{g}^{\nu\beta}}\ \right]  \rightarrow $ 
\begin{equation}
A\left[\,1+2\kappa^2 \epsilon \left[\, X
+\frac{N}{2}\sqrt{-F_{\mu \nu }^{a}F^{a}_{\alpha\beta}\bar{g}^{\mu\alpha}\bar{g}^{\nu\beta}}\ \right]\ \right] \left[\,X
+\frac{N}{2}\sqrt{-F_{\mu \nu }^{a}F^{a}_{\alpha\beta}\bar{g}^{\mu\alpha}\bar{g}^{\nu\beta}}\ \right]
\end{equation}
where the constants $A$, $B$ and $C$ are given by, $A = \dfrac{f_2}{f_2 + \kappa^2\epsilon f_1^2}$, $B = \dfrac{\epsilon \kappa^2}{4}A$ and
$C   =\dfrac{f_1^2}{4f_2}\,A$.
Therefore, the resulting dynamics outside the bag, for $\phi \rightarrow +\infty$ will be described by the effective action (expressing $B$ in terms of $A$),
\begin{equation}\label{eq:eg1outsidebag}
S_{eff, out}=\int\sqrt{-\bar{g}}d^{4}x\left[-\frac{1}{\kappa}\bar{R}(\bar{g})
+p_{out}\left(\phi,X, F\right)\right] 
\end{equation}
\[ \nonumber
p_{out}\left(\phi,X, F\right) = AX+ A \frac{N}{2}\sqrt{-F_{\mu \nu }^{a}F^{a}_{\alpha\beta}\bar{g}^{\mu\alpha}\bar{g}^{\nu\beta}} 
- (1 + N^2\epsilon \kappa^2A)\frac{1}{4}F_{\mu \nu }^{a}F^{a}_{\alpha\beta}\bar{g}^{\mu\alpha}\bar{g}^{\nu\beta}
\]
\begin{equation}\label{eq:eg1poutsidebag2}
+ AN\epsilon \kappa^2 X\sqrt{-F_{\mu \nu }^{a}F^{a}_{\alpha\beta}\bar{g}^{\mu\alpha}\bar{g}^{\nu\beta}} + A\epsilon \kappa^2X^{2} -C
\end{equation}
Working in the case  where gravitation plays an important role, one could also think of using the approach developed here to generalize the and "hiding and confining effects"  shown in chapter \ref{Lightlike Brane Sources} and  \ref{Timelike Brane Sources}, where the confining region is an uncompactified space-time and where charges send the gauge field flux they generate completely into a "tube-like" compactified region.


\newpage
\section{Summary}

We presented a gauge field system which is of a special non-linear form containing a square-root of the Maxwell term and discussed some of the motivations for considering such a term and how it can emerge from spontaneous braking of scale invariance. Another, important, feature of this term is that it produces a QCD-like confining gauge field dynamics in flat space-time. We then considered the coupling of this term to Einstein gravity and found compactified ``tube-like" solutions and non-compact solutions with a \emph{dynamically generated} cosmological constant. 
We where able to match these two solutions via light-like and time-like branes, with a main objective of searching for charge hiding and confining behaviours in curved space-time, where the role of charged objects subject to confinement is played by charged light(time)-like branes. 

We found that topologically induced charge-confining or charge-``hiding'' effects, for example the charge hiding effect consists of flux lines, of the dielectric field $D$, flowing only into the tube-like region of the horned-particle thus a truly charged object appears neutral for an observer in the non-compact universe, take place within {\em wormhole-like} solutions (the so called Horned-Particle) with
the following special structure: One of the ``universes'' comprising the total wormhole-like space-time manifold must be a compactified ``universe'' of Levi-Civita-Bertotti-Robinson
(``tube-like'') type with geometry $\mathcal{M}_2 \times S^2$ where the two-dimensional manifold $\mathcal{M}_2$. The one (or two) outer ``universe(s)'' are non-compact spherically symmetric of the Schwarzschild-type.

Due to the presence of the ``square-root'' Maxwell term a non-zero constant vacuum electric field is generated in (any of) the outer
non-compact ``universe(s)'', however, the total flux is {\em zero} there
because of vanishing of the pertinent electric displacement field. For a light-like brane this is a direct result of matching conditions, while for time-like brane this is the only possible way that a truly charged particle can still be of finite energy. Therefore, the charged light(time)-like brane occupying the ``throat'' between the non-compact
and the compactified ``tube-like'' ``universes'' appears as electrically
neutral to an external observer in the non-compact ``universe'', thus the flux is entirely confined within the ``tube-like'' ``universe''.

For a time-like brane we find that no surfaces of infinite coordinate time redshift\footnote{Such surface of infinite redshift is in fact a horizon at $r=r_h$, so the resulting object is similar to a black hole, but not exactly, in our previous studies of wormholes constructed this way, because there is no "interior", that is $r>r_h$ everywhere in these previous studies and since $r>r_h$ everywhere, there is no possibility of "collapse" to $r=0$ } appear in the problem, leading therefore to a \emph{completely explorable} horned particle space, according to not only the traveller that goes through the horned particle space (as was the case for the light-like branes), but also to a static external observer. However, both the light-like and time-like sources require the existence of exotic matter ( which violates some of the energy conditions for a ``proper" matter) to support the wormhole-like solutions, which can be of quantum mechanical origin (for example quantum fluctuations). 

When we consider extended gravity model $F(R)=R+aR^2+bR_{\mu\nu}R^{\mu\nu}$ coupled with the non-linear Maxwell term we find the existence of 
nonsingular solutions in the form of wormholes with de Sitter asymptotics (with a dynamically generated cosmological constant term at large distances), as a direct result of the modified gravitational dynamics and the non-linear gauge filed. These wormhole solutions arise naturally from the nonlinear electromagnetic field rather than requiring exotic matter to generate  a pre-designed wormhole geometry. 

When coupled to $f(R)=R + \alpha R^2$ gravity plus scalar ``dilaton'', upon deriving the explicit form of the equivalent {\em local} ``Einstein frame'' Lagrangian action, we find several physically relevant features due to the
combined effect of the gauge field and gravity nonlinearities such as:
appearance of dynamical effective gauge couplings and {\em confinement-deconfinement} transition effect as functions of the
dilaton vacuum expectation value. In addition we find that the standard Maxwell kinetic term for the gauge field is \emph{dynamically generated} even when absent in the original
“bare” theory. And generalization of the solutions in the in the ordinary Einstein-gravity.

When considering the question of bags and confinement in the framework of a theory which uses two volume elements  $\sqrt{-{g}}d^{4}x$ and $\Phi d^{4}x$, where $\Phi $ is a metric independent density. For scale invariance a dilaton field $\phi$  is considered. 
Using the first order formalism, curvature ( $\Phi R$ and  $\sqrt{-g}R^{2}$ ) terms , gauge field term ( $\Phi\sqrt { - F_{\mu \nu }^{a}\, F^{a}_{\alpha\beta}g^{\mu\alpha}g^{\nu\beta}}$ and  $\sqrt{-g} F_{\mu \nu }^{a}\, F^{a}_{\alpha\beta}g^{\mu\alpha}g^{\nu\beta}$ ) and dilaton kinetic terms are introduced in a conformally invariant way. Exponential potentials for the dilaton break down (softly) the conformal invariance down to global scale invariance, which also suffers s.s.b. after integrating the equations of motion. The model has a well defined flat space limit.  As a result of the s.s.b. of scale invariance phases with different vacuum energy density appear. Inside the bags the gauge dynamics is normal, that is non confining, while for the outside, the gauge field dynamics is confining.


\end{document}